\pdfoutput=1
\documentclass[aps,10pt,bibnotes,prb,amsmath,citeautoscript,twocolumn,
               preprintnumbers,amssymb,amsmath,floatfix,
               letterpaper]{revtex4-1}
\usepackage[bookmarks=false]{hyperref}
\usepackage{graphicx} 
\usepackage[protrusion=true,expansion=true]{microtype}
\begin{document}
%
\title{Structure and properties of functional oxide thin films: Insights from 
electronic-structure calculations}
  \author{James M.\ Rondinelli}
  \email{rondinelli@anl.gov}
  \affiliation{%
  X-Ray Science Division, Argonne National Laboratory, 
  Argonne, Illinois 60439, USA}
  \author{Nicola A.\ Spaldin}
  \email{nicola.spaldin@mat.ethz.ch}
  \affiliation{Department of Materials, ETH Zurich, Wolfgang-Pauli-Strasse 27,
8093-Z\"{u}rich, Switzerland}

\date{\today}
%

\begin{abstract}
\sloppy
The confluence of state-of-the-art electronic-structure computations 
and modern synthetic materials growth techniques is proving indispensable in the 
search for and discovery of new functionalities in oxide thin films and 
heterostructures.
Here, we review the recent contributions of electronic-structure calculations
to predicting, understanding, and discovering new materials physics in thin-film 
perovskite oxides.
We show that such calculations can accurately predict both structure and 
properties in advance of film synthesis, thereby guiding the search for
materials combinations with specific targeted functionalities.
In addition, because they can isolate and decouple the effects of various 
parameters which unavoidably occur simultaneously in an experiment -- such 
as epitaxial strain, interfacial chemistry and defect profiles -- they are 
able to provide new fundamental knowledge about the underlying physics.
We conclude by outlining the limitations of current computational techniques,
as well as some important open questions that we hope will motivate further 
methodological developments in the field. 
\end{abstract}

\maketitle
\begin{figure}
\centering
\includegraphics[width=0.40\textwidth]{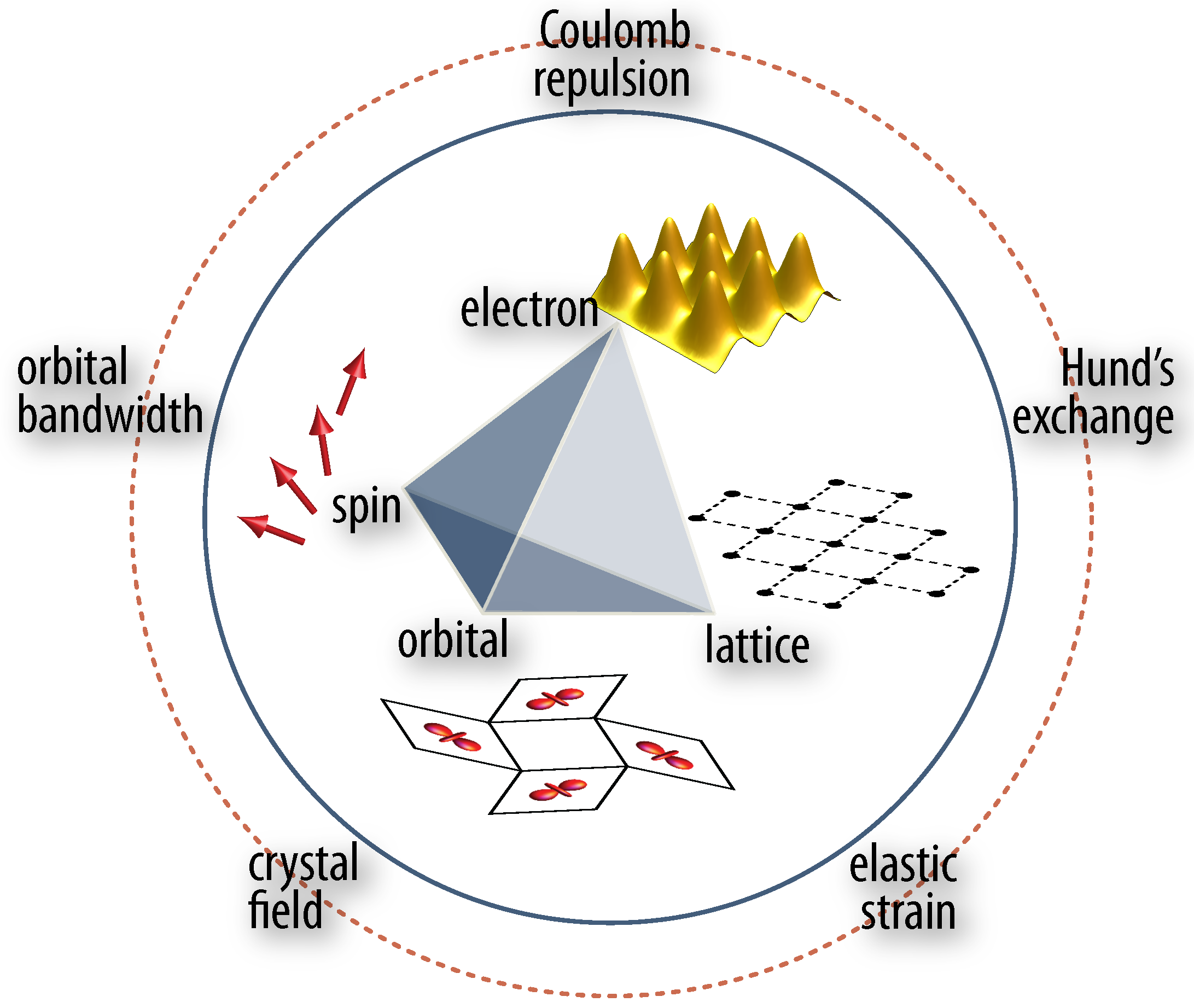}
\caption{\label{fig:couplings}
The energies of the various interactions listed around the circle are all of 
similar magnitude and the resulting competition between them leads to strong 
couplings between the electron, spin, lattice and orbital orderings illustrated
at the vertices of the tetrahedron. The couplings in turn give rise to the diverse 
functionalities of transition metal oxides, such as ferroelectricity, colossal 
magnetoresistance, and superconductivity.}
\end{figure}

\sloppy
Transition metal oxides exhibit the desirable combination
of high electronic polarizability, originating in the chemistry
of the transition metal-oxygen bonds, and strong electron correlations, 
from the localized and interacting transition metal $d$ electrons.
As a result of this combination, the energetics of various interactions 
-- such as Coulomb repulsion, strain, orbital bandwidths and Hund's
exchange -- tend to be of similar magnitude.
While in ``conventional'' materials, such as semiconductors or metals,
one of these energy scales dominates and determines the macroscopic properties, 
in transition metal oxides they compete, leading to strong 
lattice--electron, electron--spin, and spin--orbit couplings
(Figure \ref{fig:couplings}).  
The resulting ground states tend to have multiple low energy
competing phases and in turn exhibit enhanced susceptibilities 
to small external perturbations \cite{Dagotto:2005}. 
Formation of $AB$O$_3$  perovskite oxides in thin-film form
affords an additional parameter for controlling the delicate balance 
among the interactions to produce unique collective phenomena;
indeed, drastic changes in properties are reported for thin-film
oxides, such as the appearance of magnetism in otherwise non-magnetic
materials \cite{Fuchs_et_al:2007} or the activation of improper phase 
transitions.\cite{Bousquet/Ghosez_et_al:2008}
In addition, thin films provide an appropriate architecture 
for electric-field-tunable electronic, magnetic, and structural 
phase transitions, and ultimately are suitable for 
technological device integration \cite{Mannhart/Schlom:2010}.
Despite the experimental progress in achieving high quality coherent 
perovskite oxide thin films 
\cite{Schlom/Triscone_et_al:2007} and
heterostructures \cite{Nakagawa/Muller/Hwang:2006,Reiner/Walker/Ahn:2009},
there are no general rules for predicting 
the electrical, magnetic, or optical responses at oxide heterointerfaces 
given the known properties of the bulk constituents.\cite{Zubko/Triscone_et_al:2011}
This complication is due in part to our limited knowledge of the structure
of oxide thin films -- in particular, oxygen positions are highly non-trivial to
determine using standard diffraction techniques. 
In addition, the closely competing energy scales which lead to the desired
novel functionalities, in turn cause the properties to be strongly dependent 
on small changes in atomic structure and therefore hard to predict.
Thus, while oxide thin films have the potential to revolutionize the electronics 
industry through, for example, next generation {\it Mottronic}
devices \cite{Hormoz/Ramanathan:2010,Takagi/Hwang:2010,Wu/Singh:2005,Chang/Esaki:1971}, 
or could provide efficient alternatives for our growing energy needs
\cite{Shao/Haile:2004,Goodenough/Kim:2010,Kang/Ceder:2009,Voorhoeve/Gallagher:1977,Kim/Qi/Li_et_al:2010},
their adoption in practical devices has been slow. 
This is unlikely to change until a detailed microscopic understanding of the 
atomic and electronic structures in oxide thin films is developed.
This review addresses the role of electronic-structure calculations based
on density functional theory (DFT) \cite{Hohenberg/Kohn:1964,Kohn/Sham:1965} 
in confronting the complex theoretical challenge posed by oxide thin films and
heterostructures. 
Although semi-classical phenomenological models and 
well-developed theories for correlated electronic states 
have been used to describe some oxide-oxide interfaces \cite{Takashi/Nagaosa:2005,Yunoki/Dagotto/Fujimori:2007,Harrison/Grant_et_al:1978,Nakagawa/Muller/Hwang:2006,Eskes/Meinders/Sawatzky:1991,Eskes/Sawatzky:1991,Okamoto/Millis:2004a,Okamoto/Millis:2004b}, there remains no consensus as to which of 
these models (if any) is most appropriate for a general description 
of the electronic structure of oxide heterointerfaces.
A particular deficiency of methods that rely on model Hamiltonians
is that one of the energy scales in Fig.\ \ref{fig:couplings} is assumed 
to be dominant, and the delicate interplay between multiple competing 
interactions is difficult to capture. Also the atomic structure of the 
interface layers must be determined (or assumed) for input into the 
calculation.
In contrast, DFT-based techniques include all of the quantum mechanical 
interactions described by these models {\it and} the atomic structure on 
equal footing, provided that a suitable exchange-correlation potential is 
available. 
Consequently, they are able to directly explore the fundamental physics\cite{Olson:1997,Ceder:1998,Phillpot/Sinnott:2009}
of oxide heterointerfaces without {\it a priori} assumptions about which
interactions or structural distortions dominate the behavior.
As an example, DFT calculations have identified the critical role of 
strain-induced tetragonal distortions in coherent thin films that 
led to dramatic enhancements in heteroepitaxial ferroelectricity 
\cite{Choi/Schlom/Eom:2004,Haeni/Schlom:2004}, superconductivity 
\cite{Locquet/VanTendeloo_et_al:1998} or spin-phonon coupling 
\cite{Schlom/Fennie_etal:2010} depending on the material chemistry. 
It is unlikely that the importance of the tetragonal distortion
would have been identified if it had been required as an input to
the calculation rather than obtained as an output.
In this Review, we survey the current capabilities of state-of-the-art 
electronic structure approaches, and review their application to 
predicting and understanding how strain, coherency and interfacial chemistry
combine or compete to modify the properties of oxides in thin films and
heterostructures. 
We conclude by suggesting future research directions and open questions 
that electronic-structure calculations could assist in resolving.

\section{Background: Structural distortions in perovskite oxides}
\begin{figure}
\centering
\includegraphics[width=0.48\textwidth]{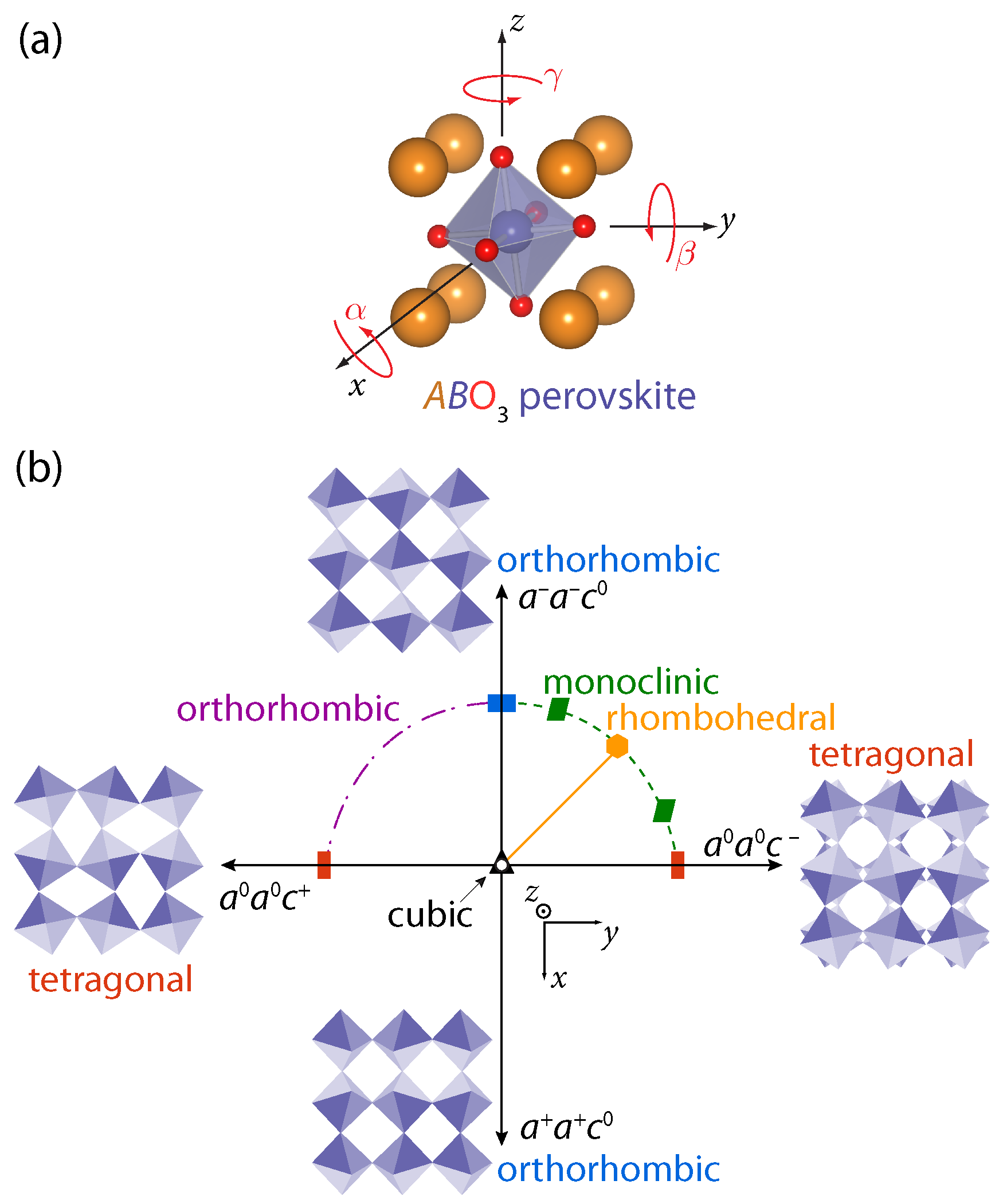}
\caption{\label{fig:perovskite_space}
Octahedral rotation phase space in perovskite transition metal oxides. 
(a) Rotations of the octahedra can be decomposed about orthogonal axes 
which intersect at the transition metal center.
(b) Representative octahedral tilt patterns described in the text:
$a^0a^0c^+$ and
$a^0a^0c^-$ correspond to in-phase ($c^+$) and out-of-phase ($c^-$) rotations of 
the octahedra about the $z$-axis, respectively, while
$a^+a^+c^0$ and 
$a^-a^-c^0$ produce similar rotations of the octahedra in the $xy$-plane.
The relative rotation of adjacent octahedra from one layer to the next 
is clearly seen for the $a^0a^0c^-$ tilt pattern (far right).
}
\end{figure}

Before beginning our discussion of the structures of thin film
perovskite oxides, we briefly review the common structural distortions 
that occur in bulk perovskites. The detailed structural distortions
that are adopted by perovskites are highly significant because they
have a profound influence on the electronic properties.

The ideal $AB$O$_3$ perovskite structure is simple cubic, 
with space group $Pm\bar{3}m$ [Figure \ref{fig:perovskite_space}(a)].
It consists of octahedrally coordinated $B$-site cations (usually transition 
metals) with three-dimensionally corner-connected $B$O$_6$ oxygen octahedra 
resulting in $\cdots$O$-B-$O$-B-$O$\cdots$ chains with 180$^{\circ}$
$B-$O$-B$ bond angles. 
Larger cations occupy the high symmetry positions of the 
cuboctahedral vacancies between the octahedra (the $A$-sites). 
Few $AB$O$_3$ oxides (including the prototype 
mineral perovksite, CaTiO$_3$) in fact adopt this ideal structure,
however, and in practice, most perovskites exhibit various structural 
distortions that lower the symmetry of the system from that of the cubic 
aristotype.
The most widely occurring distortions are rotations or ``tilts'' of
more-or-less rigid oxygen octahedra around one or more high symmetry axes. 
These are conveniently described
using Glazer notation\cite{Glazer:1972,Glazer:1975} in which
the tilt system is written 
as $a^\#b^\#c^\#$ where the letters specify the 
rotations about each pseudo-cubic axis (Figure \ref{fig:perovskite_space}), 
and the superscripts indicate whether adjacent octahedra rotate 
in-phase/ferrodistortively ($+$), out-of-phase/antiferrodistortively ($-$), or 
not at all ($0$).
Note that the decomposition relies on the octahedral units approximately 
maintaining their regularity, while strictly keeping their corner 
connectivity. As we will see later, this picture -- which is compatible 
with Pauling's rules \cite{Pauling:1929} for ionic compounds -- is
approximately correct in most cases. 
If two letters are the same, then the magnitude of the octahedral rotations, 
regardless of whether they are in- or out-of-phase, are equal.
A common misconception is that the same letters in this 
notation imply that the corresponding lattice parameters of the crystal 
are identical, which is in fact not the case---it indicates that the nearest 
neighbor transition metal distances along that direction are 
equivalent.
Since the octahedra are connected in three dimensions, a rotation or 
tilt in one direction restricts the allowed tilts and rotations in other
directions. 
In fact only 23 tilt systems can be obtained, belonging 
to 15 unique space groups \cite{Howard/Stokes:1998}; we show examples
of some common types in Figure \ref{fig:perovskite_space}, and how the 
combination of tilt systems leads to symmetry lowering of the cubic 
Bravais lattice. 
In addition to these octahedral rotations which are driven largely by 
geometric and electrostatic considerations,
\cite{Woodward:1997a,Woodward:1997b,Stokes/Howard:2005,Garcia-Fernandez_et_al:2010} 
electronically-driven distortions, particularly those caused
by the 
first- \cite{Goodenough:1998,Salamon/Marcelo:2001} and 
second-order Jahn-Teller effects 
\cite{Burdett:1981,Pearson:1983,Bersuker:2001} are 
important in determining a perovskite's structure.
First-order Jahn-Teller distortions occur when an electronic degeneracy  
usually associated with the $d$-electrons on the $B$-site cation can be 
removed by an appropriate structural distortion.
This typically manifests as an elongation of some $B$--O bonds and a 
shortening of others. 
The associated arrangement of the elongations -- called a  
{\it cooperative Jahn-Teller distortion}, or an {\it orbital 
ordering} -- determines the resulting symmetry of the system. 
Finally, relative displacements of cations and ions that result
in polar ferroelectric distortions further lower the crystal
symmetry into a polar space group.\cite{Stokes/Kisi_et_al:2002}
Since these distortions can 
be described to second order in perturbation theory they are 
often refered to as second-order Jahn-Teller effects 
\cite{Rondinelli/Eidelson/Spaldin:2009}. 
\section{Strain and interface engineering in thin film perovskites}
\begin{figure}[b]
\centering
\includegraphics[width=0.48\textwidth]{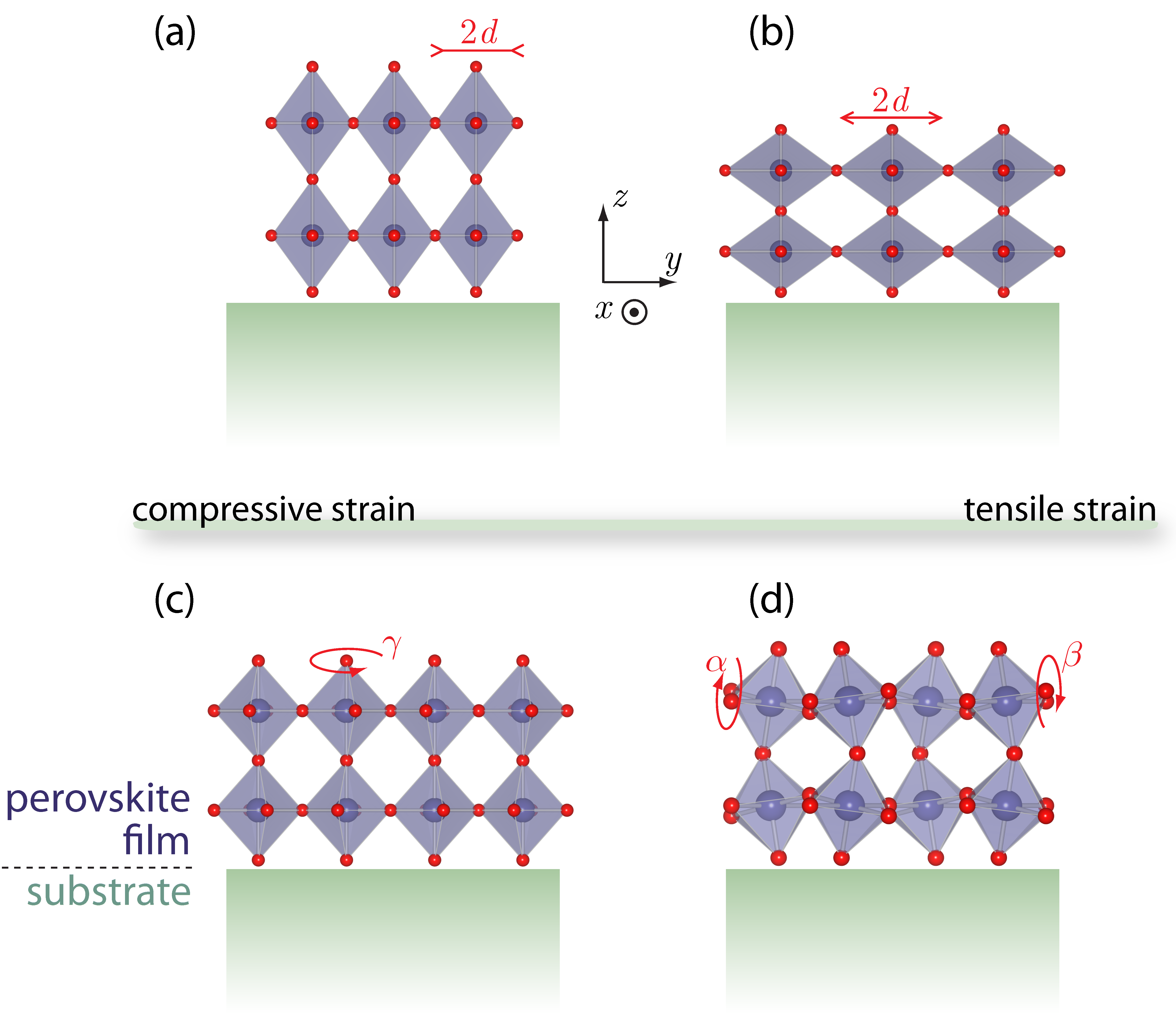}\vspace{-8pt}
\caption{\label{fig:strain_film}
In coherently strained perovskite films, the $B$O$_6$ octahedra can distort 
through contraction (a) or elongation (b) of the equatorial $B$--O bond lengths 
$d$ due to compressive or tensile strain, respectively.
Simultaneously or alternatively, the octahedra can accommodate the 
substrate-induced change of the in-plane lattice parameters by rotation 
perpendicular to the substrate as in (c), and/or about an axis parallel 
to the substrate plane (d).
}
\end{figure}
One of the primary routes to engineering the properties of a perovskite
oxide in a thin film is to leverage the elastic strain 
energy imposed by the constraint that a coherently grown film and its 
substrate have the same in-plane lattice parameters.\cite{Schlom/Triscone_et_al:2007,Martin/Chu/Ramesh:2010}
Appropriate choice of the mismatch between the lattice constants of the
substrate and the film, as well as their relative orientations, can be
used to impose a specific amount of strain on a film by the substrate.
While it is widely believed that the strain acts by imposing a new
in-plane lattice constant on the film, exactly how that change in
lattice constant is accommodated is unclear and difficult to determine
experimentally. 
We illustrate this point in Figure~\ref{fig:strain_film}. 
One possibility, shown in (a) and (b), is that the change in
in-plane lattice parameter is accommodated entirely by a change in the
in-plane metal-oxygen bond lengths. 
In panels (c) and (d) we show the other limit: the lattice mismatch 
is accommodated by a change in magnitude (or type) of the tilt patterns 
through rigid rotations of the oxygen octahedra, and  
$B-$O distances remain unchanged. 
Clearly the two responses will have drastically different effects on
the functionalities of the film. For example, changes in the $B-$O 
bond length will affect the magnitude and symmetry of the crystal field 
splitting, whereas changes in $B-$O$-B$ bond angles determine the strength
and the sign of magnetic superexchange interactions.\cite{Goodenough:1955,Kanamori:1965,Anderson:1950} 
Note that, in these simple cartoons the positions of the $A$ 
and $B$ cations are identical in the two limiting cases. 
Since, as we mentioned previously, oxygen positions are difficult to 
determine experimentally, it is difficult to distinguish between 
the two strain-accommodation limits from experiments that yield
only the cation positions
\cite{He/Xi_et_al:2004,Xie/Woicik_et_al:2008,Hoppler/Willmott:2008,Jia/Urban_et_al:2009,May/Rondinelli:2010,Vailionis/Koster_et_al:2011}.
We will see later that in most 
practical cases the actual response is intermediate between these two limits.
Another possible film response to the new substrate-enforced 
lattice parameters is that the film changes its equilibrium stoichiometry
or defect concentration.
Oxygen vacancies are a particularly common point defect in perovskite oxides 
and it is well established that materials with larger concentrations of 
oxygen vacancies have larger lattice constants.\cite{Brooks/Schlom_et_al:2009}
Since imposition of different strains requires growth on different
substrates, and associated changes in growth parameters, it is once
again difficult to establish experimentally whether changes in defect
concentrations are an intrinsic thermodynamic response to strain, or
arise from extrinsic factors during processing.
In addition to the change in the in-plane lattice parameter associated
with coherent growth on a substrate, the details of the interfacial
chemistry and structure are also likely to influence the properties 
of the film.
Here possible effects include propagation of a tilt pattern associated
with the substrate into the film,\cite{Rondinelli/Spaldin:2010b,He_Pennycook:2010} 
chemical bonding across the interface,\cite{Stengel/Spaldin:2006,Stengel/Vanderbilt/Spaldin:2009a} 
and/or interfacial electrostatics.\cite{Stengel/Spaldin:2007,Stengel/Vanderbilt/Spaldin:2009b,Pentcheva/Pickett:2010}
In the following sections, we discuss how first-principles calculations 
can identify the {\it microscopic} origins for the {\it macroscopic} material 
behavior by decoupling the effects of bi-axial strain, symmetry, 
and chemical bonding across a perovskite oxide substrate/film interface.
\section{Simulation of thin film effects -- how do the calculations actually work?}

\begin{figure*} 
\centering
\includegraphics[width=0.93\textwidth]{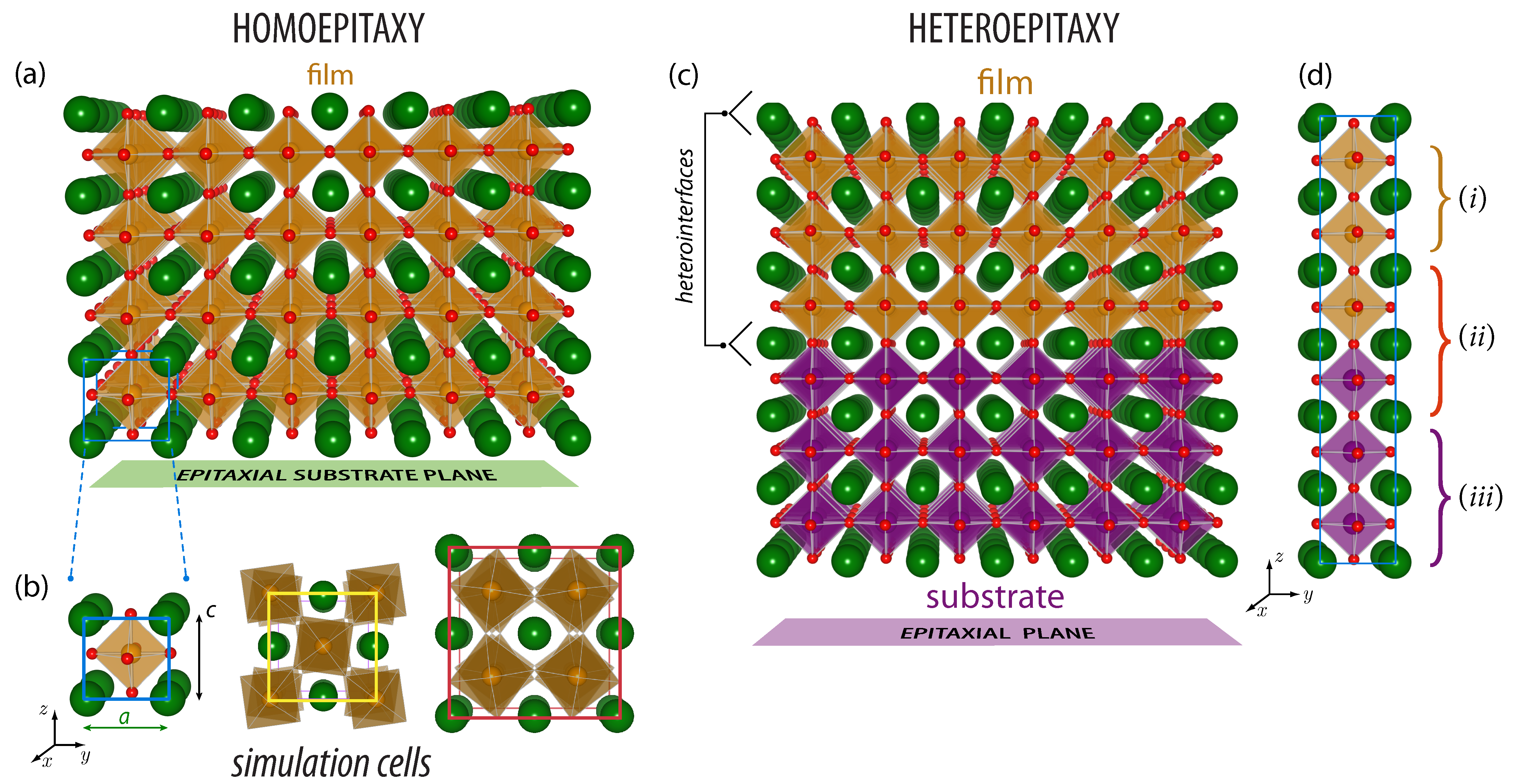}
\caption{%
Illustration of the homoepitaxial and heteroepitaxial strain models.
Possible choices for the fundamental unit cell used to represent the homoepitaxial 
film in a periodic boundary condition DFT-calculation are shown in (b). The left
cartoon shows a primitive 5-atom perovskite unit cell, and $\sqrt{2}\times\sqrt{2}\times2$ 
and $2\times2\times2$ supercells are shown center and right. 
The particular cell size is selected in order to accommodate arrangements of 
various structural and electronic internal degrees freedom.
In the heteroepitaxial strain calculations, the substrate and film are both 
present in the primitive heterostructure cell (d).
}
\label{fig:dft_tech}
\end{figure*}
In this section we describe the practicalities of how density functional
calculations for oxide thin films are carried out, with a particular focus
on their unique capability of decoupling the various competing effects that
can influence a film's behavior.
We begin with a brief review of the density functional formalism and 
then show how such calculations -- through proper choice of simulation 
cells and elastic or electric boundary conditions -- can disentangle the 
role that epitaxial strain and interface chemistry have on the macroscopic properties of perovskite oxide heterostructures.

\subsection{Density Functional Theory}
{\noindent \bf Formalism.$\quad$}
Within the density functional framework, the ground state properties of a material 
are obtained through calculation of its electron density, $\rho(\mathbf{r})$, which 
uniquely defines the energy, $E$, of the system\cite{Hohenberg/Kohn:1964}: 
\begin{equation}
\label{eq:etot}
E[\rho(\mathbf{r})] = \mathcal{F}[\rho(\mathbf{r})] + 
	\int{\!d\mathbf{r}\, V_{ext}(\mathbf{r})\rho(\mathbf{r})} \, .
\end{equation}
Here $\mathcal{F}[\rho(\mathbf{r})]$ is a universal functional describing the internal
quantum mechanical interactions of the electrons, and 
$\int{\!d\mathbf{r}\,V_{ext}(\mathbf{r})\rho(\mathbf{r})}$ is the external potential acting on the electrons
from the nuclei and any external fields.
The electronic ground state is found from
Eqn.~\ref{eq:etot} by finding the energy density that minimizes the 
total energy. This is formally equivalent to solving the many-body Schr\"{o}dinger equation 
for the fully interacting electronic system. 
However, an exact analytical form for $\mathcal{F}[\rho(\mathbf{r})]$ is not available
and in practice the system is approximated by non-interacting electrons experiencing 
an effective potential that is formulated to capture the important many-electron 
effects\cite{Kohn/Sham:1965}. 
This effective potential contains the non-interacting kinetic energy of the effective
single-particle states, the classical Coulombic interactions between them, and a 
quantum mechanical ``exchange--correlation'' energy term, $V_{xc}$, that approximates
the remaining quantum mechanical electron-electron interactions. 
Choice of the exact form of $V_{xc}$ is particularly important in calculations for
transition metal oxide heterostructures, because the strongly localized transition
metal $3d$ and oxygen $2p$ electrons, combined with localization effects introduced
by size quantization, lead to explicit and strong electron correlations. 
For example, the widely used local spin density approximation (LSDA), which uses a
parameterized form\cite{Ceperley/Alder:1980} of the 
calculated exchange--correlation energy of the uniform 
electron gas, is often inappropriate since transition metal oxides show 
large density variations.  
Better approximations are the generalized gradient approximation (GGA) 
\cite{Perdew/Wang:1986,Perdew/Burke/Wang:1996,Perdew/Burke/Ernzerhof:1996,Wu/Cohen:2006}, 
which takes into account variations in the electron density through gradient terms,
or hybrid functionals, which combine a small amount of orbital-dependent Hartree-Fock 
exchange with local or gradient-corrected density approximations
\cite{Becke:1993,Adamo/Barone:1999,Ernzerhof/Scuseria:1999,Heyd/Scuseria/Ernzerhof:2003,Bilc/Ghosez_et_al:2008,Marsman/Kresse_et_al:2008}.
Perhaps the most successful simple extension to the LSDA is the LDA$+U$ method, in which an 
orbital-dependent energy cost for adding an additional electron to an already 
occupied manifold mimics in spirit the Hubbard $U$ Coulombic 
repulsion\cite{Anisimov/Lichtenstein:1997}.
In all cases, care must be taken that the choice of functional is appropriate for
the materials to be studied if physically meaningful results are to be obtained.\\

{\noindent \bf Extracting material properties.$\quad$}
By applying suitable optimization techniques to Eq.\ \ref{eq:etot}, the ground state 
charge density and total energy can be readily obtained for a fixed crystal structure
with a specific lattice geometry and atomic positions.
All ground state properties of the system -- such as magnetic ordering, densities
of electronic states, ferroelectric polarizations, etc. -- can then be directly obtained 
from the charge density, in some cases with external fields included explicitly in the 
Hamiltonian during the 
optimization\cite{Souza/Iniguez/Vanderbilt:2002,Stengel/Spaldin:2007,
Bousquet/Spaldin/Delaney:2011}. 
In particular, the atomic forces and stresses can be computed, and the ionic positions and
cell parameters adjusted so that the forces and stresses are reduced to zero. 
This ``relaxation'' yields the lowest energy, ground-state structure.
Since in general the relaxation is a complex, multi-variable problem with many possible
local minima,\cite{Rondinelli/Marks:2007} it is usually achieved in practice by comparing trial 
structures with different crystal symmetries, obtained by freezing in combinations 
of the unstable phonons calculated for a high symmetry reference phase. 
The lattice phonon frequencies and eigenvectors can in turn be used in the 
interpretation of Raman and infrared spectroscopies as well as for the study of 
structural phase transitions.

\subsection{Homoepitaxial strain.}
We call the first type of simulation that is commonly employed 
the {\it homoepitaxial} strain approach. In homoepitaxial strain
calculations, we, in fact, model a crystal using a simulation cell
that is periodic in all 
three dimensions and subject it to a strain parallel to a chosen 
lattice plane by varying the lattice parameters away from their
equilibrium values
in that plane [Figure~\ref{fig:dft_tech}(a)]. 
Such simulations then allow the intrinsic role of epitaxial strain 
imposed by lattice matching with a substrate to be determined separately 
from any other effects associated with the presence of the interface; this is 
clearly unfeasible experimentally. 
Often bi-axial strain is applied, and in this case the strain is given by 
\begin{equation}
\epsilon = \frac{(a-a_0)}{a_0}\, , 
\end{equation}
where $a$ is the in-plane lattice parameter imposed
on the fundamental unit cell  [Figure~\ref{fig:dft_tech}(b)] 
in the homoepitaxial strain 
calculation, and $a_0$ is the calculated equilibrium lattice parameter of the bulk 
material. Different compressive ($a<a_0$) and tensile ($a>a_0$) strain states 
are obtained by varying $a$. 
Uniaxial strain can be applied by varying only one of the in-plane lattice
parameters, and anisotropic biaxial strain by varying the two in-plane parameters
by different amounts.
In all cases the out-of-plane $c$ lattice parameter and atomic positions are
relaxed to minimize the stress on the unit cell and forces on the ions, respectively. \\

{\noindent \bf Choice of simulation cell.$\quad$}
Care should be taken to select a simulation cell that allows exploration
of all likely tilt, rotation, and orbital ordering  patterns that 
might be induced by the epitaxial strain constraints. 
As we will see below in the results section, the tilt pattern in the 
film is often different from that in the bulk parent phase, and 
the default primitive simulation cell that correctly accounts for the 
bulk system's structure might not have sufficient 
flexibility to accommodate the lowest energy structure in the film.
In such cases, a \textit{supercell} is constructed from multiplication 
(and at times rotation) of the primitive cell's lattice vectors.
Such examples of supercells that can be used to perform homoepitaxial 
strain calculations are shown in Fig.\ \ref{fig:dft_tech}(b). Each of these 
simulation cells are an integer multiple of formula units (f.u.) of the primitive 
5-atom perovskite building block.
A drawback with the larger cells, however, is that they are more computationally 
expensive.
The importance of selecting the size of the supercell is best illustrated through 
a simple example.
The bulk low-temperature structure of SrTiO$_3$ has an $a^0a^0c^-$ 
tilt system, which requires a supercell with a
$\sqrt{2}\times\sqrt{2}\times2$ (20 atom) larger periodicity than that of the ideal
5-atom perovskite [Fig.\ \ref{fig:dft_tech}(b,center)].
If epitaxial or uni-axial strain were to stabilize the  mixed-tilt 
$a^0b^+c^-$ pattern that is found in SrZrO$_3$, however, a 2$\times$2$\times$2 
(40 atom) supercell would be needed.
The smaller unit cell would fail to find the global ground state structure and 
the results could be misleading.
We survey in the Results section, how simulation cell choice is crucial 
to identifying strain-induced changes to the oxygen octahedral rotations in 
perovskites.\\

{\noindent \bf Symmetry consequences.$\quad$}
Even at the homoepitaxial strain level, without explicit inclusion of 
the substrate in the calculation, the imposition of specific in-plane 
lattice parameters often lowers the lattice symmetry of the film. 
Here, we look briefly at why this occurs and discuss the implications of 
the symmetry change on both the practicalities of the calculations and 
the macroscopic film properties.
\begin{figure}
\centering
\includegraphics[width=0.48\textwidth]{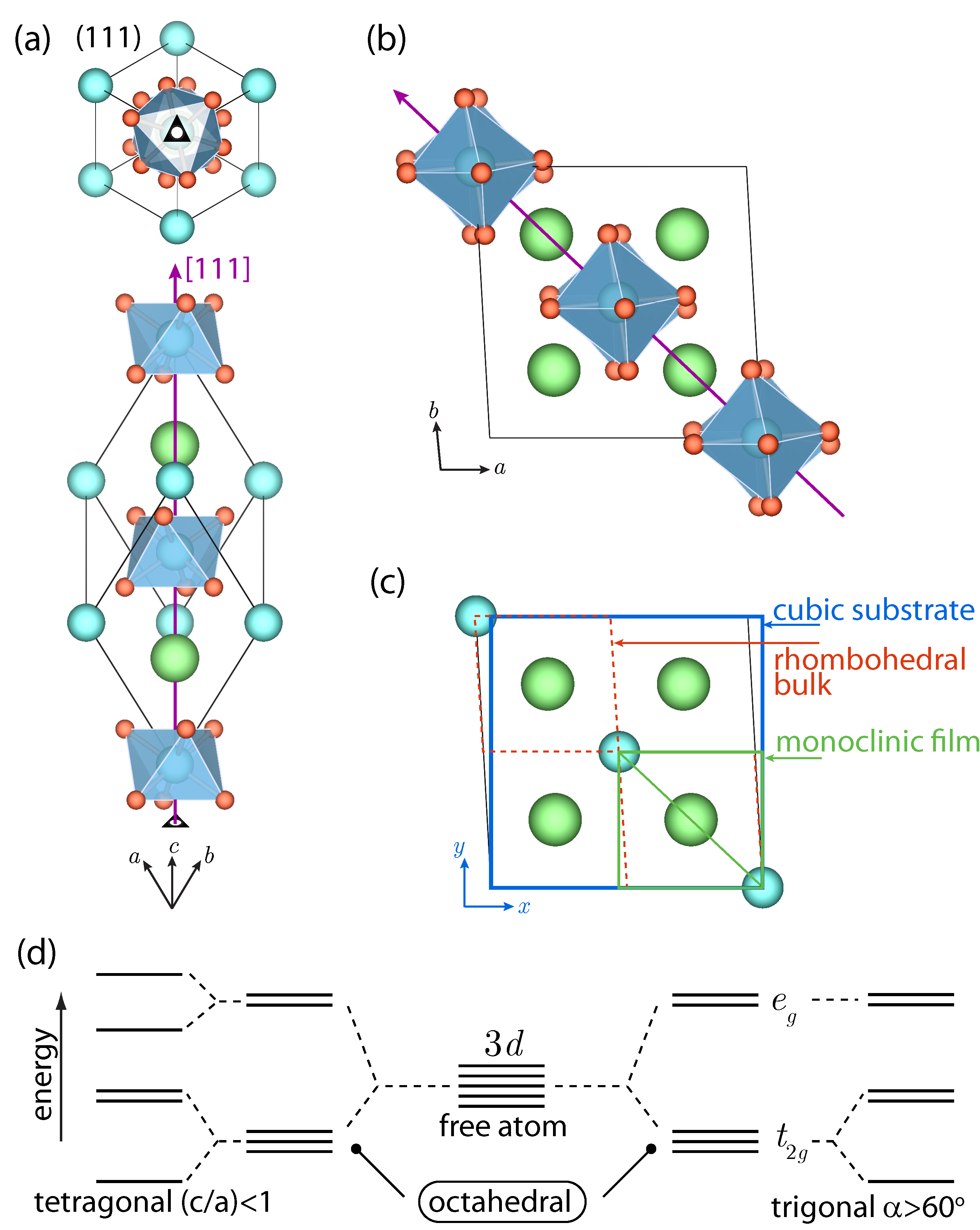}
\caption[Orientation of a rhomboherdal perosvkite film on a cubic substrate]{\label{fig:rho:twin} 
(a) Structure of a rhombohedral perovskite with symmetry $R\bar{3}c$ exhibiting 
$a^-a^-a^-$ octahedral rotations about the three-fold [111] axis (red arrow).
(b) Projection of the three dimensional structure  shown in (a) into the 
$ab$ epitaxial plane. 
In (c) we show the same projection without the oxygen atoms for clarity. The blue square
shows how forcing the rhombohedral cell to match with a cubic substrate reduces the
symmetry in this case to monoclinic. 
The effect of different strain symmetries in the perovskite film 
on the 3$d$ one-electron orbital level splittings is shown in (d).}
\end{figure}

We first describe the possible symmetry modifications with the  
simplest possible example: A material which in the bulk has 
cubic symmetry (space group $Pm\bar{3}m$), and which is grown
on a (001) interfacial plane of a substrate also with 
cubic symmetry. An example of such a material would be 
an ideal cubic perovskite, with no symmetry-lowering rotations
of the oxygen octahedra, Jahn-Teller or ferroelectric distortions. 
When biaxial strain is imposed ($a \ne a_0$), it is clear that 
the space group symmetry of the material is immediately reduced 
to tetragonal ($P4/mmm$), with the three-fold axes along the body 
diagonal of the unit cell removed, but the four-fold and mirror 
symmetry elements preserved.
Note that of course, the symmetry can be further lowered due to internal 
structural distortions. For example, a common effect of in-plane compressive
strain on perovskite dielectrics is to induce a ferroelectric polarization along the 
$c$-axis.\cite{Dawber/Rabe/Scott:2005} 
In this latter case, an additional mirror plane is lost and the space group 
symmetry is reduced further to $P4mm$.
We next illustrate a more complex situation in which a rhombohedral perovskite -- 
characterized by octahedral rotations around the three fold axis along 
the pseudo-cubic [111]-direction [Figure \ref{fig:rho:twin}(a)] -- 
is placed on an (001)-oriented cubic substrate.
The anti-phase rotation of the $B$O$_6$ octahedra about the [111] axis 
leads to a shear distortion in the interaxial cell angles. This can be seen
in Figure~\ref{fig:rho:twin}(b), in which we have projected the
three-dimensional structure onto the two-dimensional $ab$-plane; the angle
between the $a$ and $b$ lattice vectors clearly deviates from 90$^{\circ}$. 
Imposition of a coherent epitaxial constraint onto a cubic (100) substrate,
however, forces the in-plane inter-axial angle to be 90$^\circ$, as shown
in Figure~\ref{fig:rho:twin}(c). 
Since the epitaxial constraint does not impose restrictions on the out-of-plane 
angles, the formerly three-fold axis is reduced to a $\frac{2}{m}$ operation resulting
in monoclinic symmetry.\cite{Daumont/Noheda_et_al:2010}
In Table \ref{tab:lattices}, we list the symmetry reductions that occur
for other common lattice types on a range of common substrate symmetries 
and orientations.

\begingroup
\squeezetable
\begin{table}
\caption{\label{tab:lattices} 
Symmetry consequences on the film lattice type due to two-dimensional 
thin film epitaxy. In all cases the orientation of the substate is given 
with reference to the pseudocubic lattice vectors of the aristotype 
perovskite phase. Note that an appropriately oriented tetragonal substrate 
will result in the same symmetry lowering as a cubic substrate with
bi-axial strain.}
\begin{ruledtabular}
\centering
\begin{tabular}{llll}%
	& \multicolumn{3}{c}{Parent film under bi-axial strain} \\ 
\cline{2-4}
Substrate		&  cubic        & rhombohedral	& orthorhombic \\
\hline
cubic (001)		& tetragonal	& monoclinic	& orthorhombic \\
cubic (110)		& orthorhombic	& triclinic	& monoclinic   \\
rhombohedral (001)	& monoclinic	& triclinic	& triclinic    \\
orthorhombic (001)	& orthorhombic	& triclinic	& orthorhombic \\
\end{tabular}
\end{ruledtabular}
\end{table}
\endgroup
The reductions in lattice symmetry discussed above of course have consequences
on the properties. 
Changes in the point symmetry of the transition metal cation site can
modify the crystal field splitting of the $d$ orbitals. 
For example, in Figure \ref{fig:rho:twin}(d) we show the effect on the $d$-orbital 
energy levels of reducing an octahedrally coordinated  transition metal ion's symmetry 
to tetragonal (left) or trigonal (right).
The former is often believed to be the experimental situation in 
ultra-thin perovskite films on a square substrate in which 
pseudomorphic growth is maintained.\cite{Schlom/Triscone_et_al:2007} 
In contrast the latter is likely for thicker films of perovskites with 
bulk rhombohedral phases in which lattice relaxations have occurred.\cite{Fuchs_et_al:2009}
Such changes in the crystal-field splitting can have dramatic effects in $d^n$ perovskite 
oxides by modfying orbital degeneracies and in turn Jahn-Teller and orbital ordering 
patterns, as well as allowing transition metal $d$-electrons 
with different orbital angular momentum to mix more (or less) strongly.\\ 
Finally, we point out that homoepitaxial strain simulations offer an exciting 
complement to experimental studies of strain effects on thin film properties 
since the substrate's lattice parameter $a$ can be tuned  continuously in the 
calculations to explore the full strain--structure--property phase space. As
a result it is possible to identify critical strain values at which phase 
transitions should occur and susceptibilities diverge.
In contrast, the experimental situation tends to be less flexible since 
the choice of substrates on which to grow thin films is limited by the 
availability of high purity single crystals, and the equilibrium volumes 
for their particular chemical compositions.
Of course homoepitaxial calculations omit explicit structural and 
chemical interactions with the substrate material, which in some cases 
might be important and even dominate the strain-induced behavior. 
We discuss how such factors are incorporated in
so-called {\it heteroepitaxial} simulations next.
\subsection{Heteroepitaxial strain simulations}
In heteroepitaxial strain simulations a two-component supercell is used,
with a second material included explicitly to model the presence of a
substrate [Figure~\ref{fig:dft_tech}(c) and (d)]. 
With both constituents included, it is possible to identify 
how the physical and electronic structure of the atoms in the 
different layers of the film ($i$), interface ($ii$), and/or substrate ($iii$) 
contribute to the properties of the heterostructure.
Then, by comparing the results of homoepitaxial strain calculations with these 
{\it heteroepitaxial} strain calculations, the role of the substrate and 
interface can be isolated. 
The first step in a heteroepitaxial strain simulation is the selection
of a unit cell that is appropriate for studying the properties of interest. 
An example is shown in Figure~\ref{fig:dft_tech}(d).
Since periodic boundary conditions are used, in practice the
calculations model a periodic superlattice with an infinite array of 
interfaces rather than a single heterointerface.
If this is indeed the experimental situation, then the repeat unit should be
chosen to most closely match the periodicity of the experimental superlattice 
within the limits of available computer resources. 
In this case, it is usually appropriate to relax all lattice parameters
and internal atomic positions to their lowest energy values. 
If instead the calculations aim to answer questions about a single component
film on a thick substrate, the supercell should be chosen to be as large as 
possible so that interactions between the interfaces are minimized. In this 
limit, the in-plane lattice constant is usually constrained to that of the substrate,
and the atoms in the middle layers of the substrate material are often fixed to their
bulk positions.
A key question to be addressed in heteroepitaxial calculations is whether
the tilt and rotation patterns found in the substrate template across
the interface into the film material, and if so, how do these distortions 
modify the macroscopic properties. 
In Table \ref{tab:tilts}, we list the octahedral tilt patterns
adopted by many widely used substrate materials at various 
temperatures. 
The propagation of these tilt patterns across interfaces has sometimes 
been invoked in the literature to explain
observed phenomena \cite{He/Xi_et_al:2004,Xie/Woicik_et_al:2008}
despite being difficult to confirm experimentally.
Density functional calculations with full or constrained structural optimizations
are a powerful tool for testing the veracity of this assumption. 
In addition, large superlattices make it possible to analyze layer-by-layer changes 
in the atomic and electronic structure to identify whether there is a 
critical thicknesses at which the film recovers its bulk tilt pattern and
other properties.
\begin{table*}
\caption{\label{tab:tilts} 
Lattice parameters, crystal structures and tilt systems of common 
substrate materials used in oxide thin film growth.
Pseudo-cubic lattice parameters are given in parentheses. For 
rhombohedral space groups $a_{pc} \approx a/\sqrt{2}$.
For orthorhombic substrates values in parentheses correspond to 
the average pseudocubic spacing ($\sqrt{a^2+b^2}/2$) 
along the [110] direction.}
\begin{ruledtabular}
\centering
\begin{tabular}{lllllc}%
Substrate		& Structure	& Temperature & Tilt System	& Lattice constants (\AA) & Reference \\
\hline
SrTiO$_3$ (STO)		& cubic (221, $Pm\bar{3}m$)  &$>105$~K & $a^0a^0a^0$	&	$a=3.905$	& \onlinecite{Unoki/Sakudo:1967}	 \\
			& tetragonal (140, $I4/mcm$) &$<105$~K & $a^0a^0c^-$	& 	$c/a=1.0056$ 	& 	\\
\hline
LaAlO$_3$(LAO)	& cubic (221, $Pm\bar{3}m$)        &$>800$~K & $a^0a^0a^0$	&$a=3.81$ &	\onlinecite{Geller/Raccah:1970} \\
			& rhombohedral (167, $R\bar{3}c$) & $<800$~K & $a^-a^-a^-$	&$a=5.36$ (3.79)  &	\onlinecite{Berkstresser_et_al:1991} \\
\hline
LSAT			& cubic (221, $Pm\bar{3}m$)        &$>150$~K & $a^0a^0a^0$	&$a=3.87$ 	& \onlinecite{Chakoumakos/Schlom_et_al:1979}	\\
			& tetragonal (140, $I4/mcm$) &$<150$~K & $a^0a^0c^-$	& 	$a=5.46,c=7.73$	& \\
\hline
LaGaO$_3$		& rhombohedral (167, $R\bar{3}c$) & $>420$~K & $a^-a^-a^-$	&$a=5.58$ (3.94)	& \onlinecite{Marti/Tandin_et_al:1994}	\\
			& orthorhombic (62, $Pnma$)	&  $<420$~K & $a^+b^-b^-$	& $a=5.49,b=5.53,c=7.78$ (3.89)	&	\onlinecite{Utke_et_al:1997} \\
\hline
DyScO$_3$	(DSO)	& orthorhombic (62, $Pnma$)	& & $a^+b^-b^-$	& 	$a=5.44,b=5.71,c=7.89$ (3.94)	&	\onlinecite{Schubert_et_al:2003}\\
\end{tabular}
\end{ruledtabular}
\end{table*}

\begin{figure}
\centering
\includegraphics[width=0.42\textwidth]{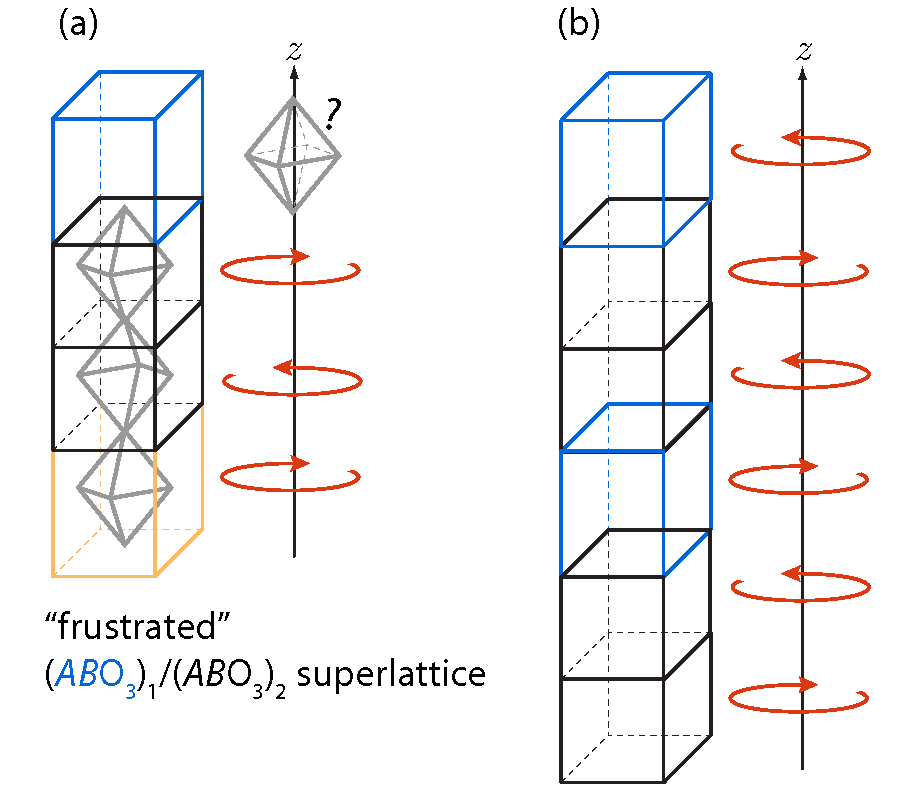}
\caption{\label{fig:tilts_frustration}
Example of a frustrated antiferrodistortive octahedral rotation system 
in an odd period superlattice (a).
The sense of the octahedral rotation is indicated by the red arrows, and due to 
the chosen periodicity, a rotation direction cannot be selected in the blue 
block that is compatible with the periodic boundary conditions.
To eliminate the artificial octahedral frustration in the superlattice,  
a supercell doubled along the $z$-direction is needed to 
be fully compatible with the $a^0a^0c^-$ rotation pattern.}
\end{figure}

%
An additional point to note in heteroepitaxial strain calculations 
is that the internal distortions of the atoms, 
primarily the rotations of the octahedra, must also be compatible with the size 
of the unit cell used to simulate the superlattice.
As discussed in the case of homoepitaxial strain, the chosen in-plane periodicity
must allow the necessary tilt and rotation patterns; the 
supercell shown in  Figure \ref{fig:dft_tech}(d), with its single perovskite
unit cell in plane, clearly prohibits this. In addition the out-of-plane
periodicity must be chosen to avoid artificial frustration of the tilt system 
(Figure~\ref{fig:tilts_frustration}).
For example, the out-of-phase $a^0a^0c^-$ tilt pattern
requires an even number of 5-atom perovskite blocks along the rotation axis to 
accommodate the full periodicity of the zone-boundary phonon mode.
If, for example, 1/2-period perovskite superlattice with only three 5-atom blocks 
is simulated in this tilt pattern, as shown 
schematically in Figure \ref{fig:tilts_frustration}(a), 
the octahedral rotation in the last block becomes frustrated: The unit cell
below it expects it to rotate clockwise, whereas that above it expects it
to rotate anti-clockwise.
A possible solution is to double the primitive heteroepitaxial cell along
the rotation axis and then perform calculations on the 
supercell depicted in Figure \ref{fig:tilts_frustration}(b).\\
\section{Results: Homoepitaxial strain}
\subsection{``Simple'' examples: Strain--octahedral tilt coupling 
in rhombohedral LaAlO$_3$ and LaNiO$_3$}
We begin our results section with a discussion of the behavior of two
ostensibly simple examples: LaAlO$_3$ and LaNiO$_3$ films strained on 
(001)-oriented cubic substrates. 
Both compounds crystallize in the rhombohedral $R\bar{3}c$ space 
group with the  $a^-a^-a^-$ tilt pattern, and in both  cases, 
the bulk ground state structure consists of only rotational 
distortions from the ideal cubic structure. 
The bulk pseudocubic lattice parameters are 3.79 
and 3.84 \AA\ respectively, resulting in Al-O-Al (Ni-O-Ni) bond
angles of 171$^\circ$ (165$^\circ$) and Al-O (Ni-O) bond lengths of 
1.90 (1.94) \AA. 
In this symmetry class, the $B$O$_6$ octahedral rotation angle 
is known to vary strongly with the shape of the Bravais lattice 
\cite{Megaw/Darlington:1975,Thomas:1996,Guo/Rondinelli:2011}. Therefore, we expect that 
changes in the degree of monoclinicity introduced by the 
substrate strain  (Table ~\ref{tab:lattices}) will have a large effect
on the internal atomic positions. 
Finally, LaAlO$_3$ is a robust wide band-gap insulator with no tendency
to ferroelectric distortion, whereas LaNiO$_3$ is metallic, with the 
additional complexity of partially filled $d$ orbitals, suggesting 
proximity to Jahn-Teller or charge disproportionation instabilities, 
as well as possible metal-insulator transitions.
\cite{Lacorre/Torrance:1992,Torrance/Niedermayer_etal:1992,Zhou/Bukowski:2000}
These factors combine to make LaAlO$_3$ and LaNiO$_3$ model systems for studying 
coupling between strain and octahedral rotations, and the influence of additional
complexities on this coupling.

\begin{figure}[b]
\centering
\includegraphics[width=0.38\textwidth]{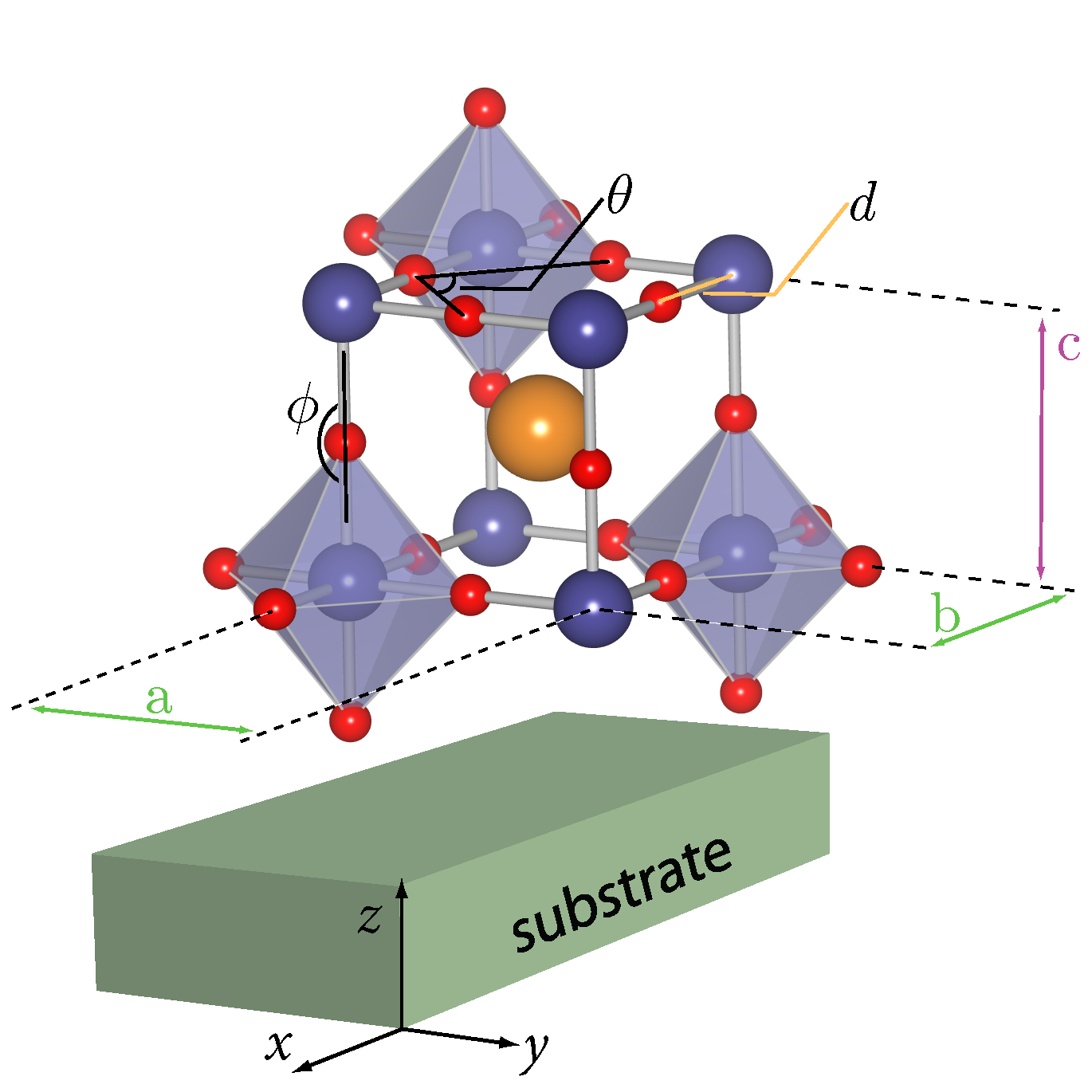}\vspace{-6pt}
\caption{\label{fig:rotations_angles_only}
Definitions for the rotation ($\sim\phi$) and tilt ($\sim\theta$) angles that are 
often used in the literature to describe substrate-induced changes to 
the Glazer octahedral tilt patterns. 
}
\end{figure}

\begin{figure*}
\centering
\includegraphics[width=0.67\textwidth]{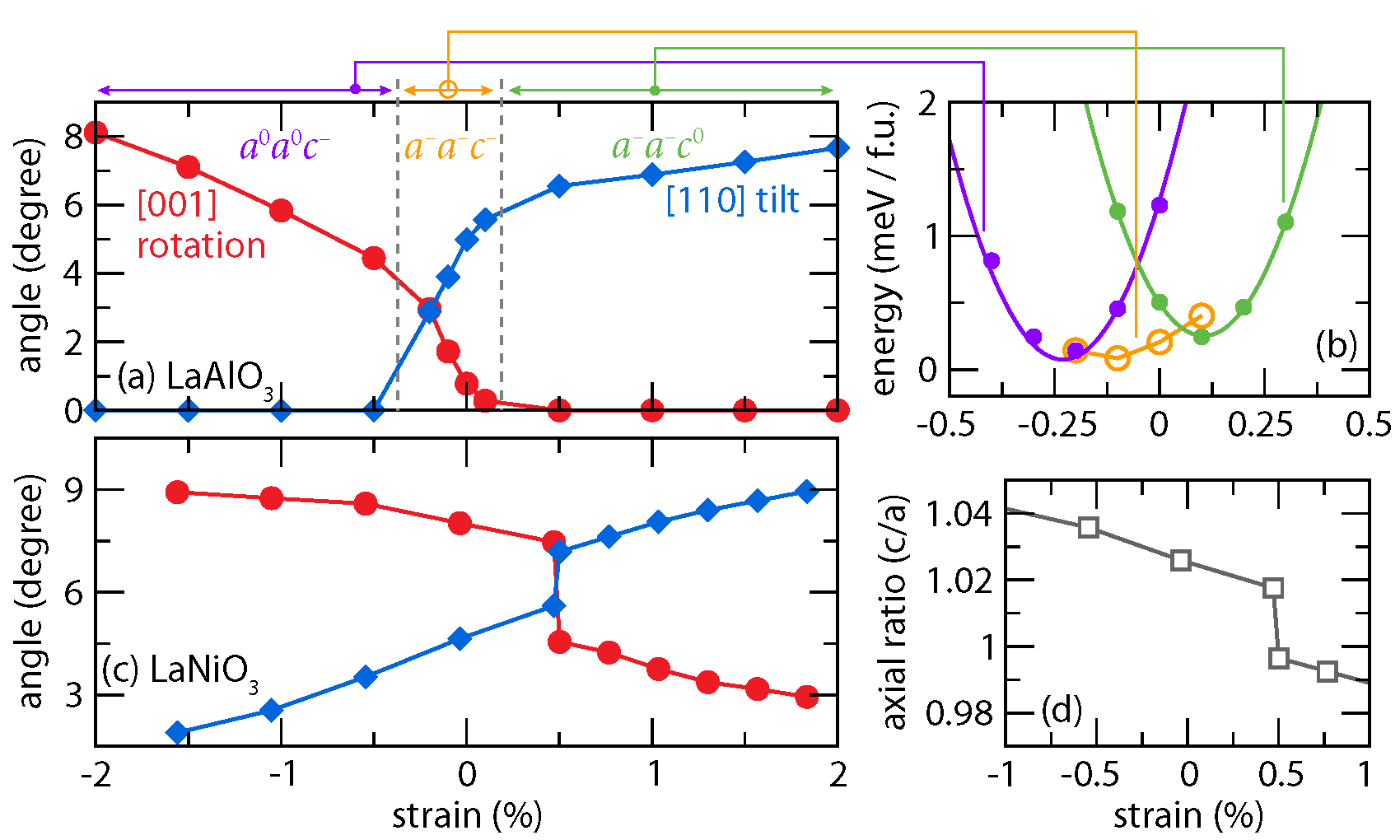}
\caption{\label{fig:homostrain_data} Zero kelvin strain--octahedral 
phase diagrams for two bulk rhombohedral perovskites.
(a) Change in the AlO$_6$ octahedral rotation angles in LaAlO$_3$ and 
(b) the energy stabilization for each phase with the corresponding tilt 
patterns.
(c) In LaNiO$_3$, the $a^-a^-c^-$ tilt pattern is found to be the ground state 
for all strains investigated. The ``sense'' of the $a^-a^-c^-$ 
octahedral rotation pattern in LaNiO$_3$ 
changes abruptly as the axial ratio approaches unity (d).
Data reproduced with permission from Refs.~\onlinecite{Hatt/Spaldin:2010} 
and~\onlinecite{May/Rondinelli:2010}, Copyright 2010, {\it American Physical Society}.}
\end{figure*}
To investigate the possible rotation patterns present in a homoepitaxial strained film, 
it is useful to first use symmetry analysis to determine the possible octahedral tilt 
systems and space groups that are compatible with symmetry of the cubic (001) substrate.
In this case these are:
$a^-a^-c^-$ (15 $C2/c$ or 14 $P2_1/c$),  $a^-a^-c^+$ (62 $Pnma$), 
$a^+a^+c^-$ (137 $P4_2/nmc$), 
$a^-a^0c^0$ (69 $Fmmm$), 
$a^-a^-c^0$ (74 $Imma$), 
$a^0a^0c^0$ (221 $Pm\bar{3}m$), 
$a^+a^0c^0$ (55 $Pbam$), 
$a^+a^+c^0$ (139 $I4/mmm$), 
$a^0a^0c^-$ (140 $I4/mcm$), $a^0a^0c^+$ (127 $P4/mbm$), and 
$a^+a^+c^+$ (71 $Immm$).
If experimental pressure or temperature phase diagrams are available, 
one can also further narrow this set to likely candidates.  
The relative energies of these structural phases are then calculated and 
compared to determine the lowest energy structure at each strain state.
Following identification of the important tilt patterns, the change in magnitude of 
the octahedral rotations about each direction for a given 
Glazer tilt system is commonly described in terms of octahedral 
``rotations'' and ``tilts.'' 
These octahedral connectivity descriptors are defined relative 
to the substrate orientation (Fig.\ \ref{fig:rotations_angles_only}): 
The tilt angle is given as 
$(180-\phi)/2$ and the rotation angle is $(90-\theta)/2$. 
In this way, the effects of substrate-strain on the octahedral $B$--O--$B$ 
bond angles can be linked to the material properties.
For a film grown on a substrate with a square net (bi-axial strain), 
the geometric relationships between Glazer's rotation angle definitions about the  
pseudo-cubic lattice vectors (Figure \ref{fig:perovskite_space}) to the 
rotation $\theta$ and tilt $\phi$ angles are as follows: 
$\theta=\gamma$ and $\phi = (\alpha + \beta) / \sqrt{2}$.
Using the octahedral symmetry guided approach, 
Hatt and Spaldin mapped out the octahedral--strain 
phase diagram for LaAlO$_3$ [Figure \ref{fig:homostrain_data}(a)], 
which we reproduce and describe here.\cite{Hatt/Spaldin:2010}
We see in Figure \ref{fig:homostrain_data}(a) that for very small strains, 
the rhombohedral-like ($a^-a^-a^-$) pattern of the bulk system persists,
albeit with a monoclinic lattice distortion from
constraining the in-plane lattice parameters and angle.
Small compressive strains of $0.2\%$, however, are able to fully suppress the 
bulk octahedral $a^-a^-a^-$ tilt pattern and change it to $a^0a^0c^-$, which has
rotations around the out-of-plane axis only. 
In this region, the lattice is tetragonal 
with 90$^\circ$ interaxial angles and space group $I4/mcm$. 
This behavior can be na{\"i}vely understood in terms of the reduction in area perpendicular
to the rotation axis in response to the in-plane compressive strain.
In contrast, under tensile strain, the $a^-a^-c^0$ pattern is stabilized, with
rotations around the [110] axis only (space group $Imma$).
This rotation pattern reduces the length of the film's unit cell along 
the out-of-plane direction (in accordance with elastic theory due to the 
tensile strain in plane) while keeping the Al--O bonds nearly 
constant.
In Figure \ref{fig:homostrain_data}(b) we show the calculated total energies of LaAlO$_3$ 
as a function of in-plane biaxial strain. We see that the energy differences between different
patterns of tilt and rotation at typical strain values are only a few meV/f.u.
In Ref. \onlinecite{Hatt/Spaldin:2010} it was also found that the energy of 
unstrained but {\it coherent} LaAlO$_3$ is higher than that of the relaxed
bulk rhombohedral structure. If all interaxial angles are fixed to be 90$^\circ$
-- as is often assumed in the literature -- 
the increase in energy is 0.6~meV/f.u., whereas when a relaxation to monoclinic
symmetry is allowed it is only 0.2 meV per formula unit higher in energy. This
suggests that the films will in practice exhibit monoclinic distortions.
Since the nature of the octahedral rotations depends on a delicate balance between 
bonding and electrostatic interactions in the solid,  we might expect the behavior
in metallic LaNiO$_3$ to differ substantially from that in insulating LaAlO$_3$. 
Therefore, we next describe the octahedral structure evolution with strain for LaNiO$_3$ 
[Figure \ref{fig:homostrain_data}(c)]. 
While the variations in tilt and rotation angles show some similarity
to the LaAlO$_3$ case -- the amount of [001] rotation is increased with compressive
strain and decreased with tensile strain, with the opposite behavior for the [110]
tilt -- in this case the monoclinic $C2/c$ structure persists over the whole
strain range. 
A recent detailed x-ray diffraction study of strained 
LaNiO$_3$ films on SrTiO$_3$ and LaAlO$_3$ substrates by 
May and co-workers \cite{May/Rondinelli:2010} yielded refined structures 
in excellent argeement with these computational results: 
neither the tilts nor the rotations are ever completely de-stabilized by the 
substrate-induce strain. 
It is clear from Fig.~\ref{fig:homostrain_data} that, while there is no change in 
symmetry for LaNiO$_3$ over the strain range investigated, there is a distinct structural transition
characterized by a discontinuity in the magnitude of the tilt and rotation angles;
this corresponds to an abrupt reorientation in the axis about which the NiO$_6$
octahedra rotate.  
This first-order phase transition is classified as isosymmetric 
since the atomic structure remains monoclinic and the full symmetry and 
Wyckoff positions of the $C2/c$ space group are retained \cite{Christy:1995}.  
(We note the authors of Ref.\ \onlinecite{May/Rondinelli:2010} predict 
an additional phase of lower symmetry ($P2_1/c$) in LaNiO$_3$ films under tensile strain  
which shows a small charge disproportionation (CDP) but with negligible 
changes in the rotation angles. It is approximately 
2 meV lower in energy and would result in a ``normal'', non-isosymmetric
transition.)
Interestingly, the isosymmetric phase transition occurs where the axial ratio of the 
crystal approaches unity [Figure \ref{fig:homostrain_data}(d)].
This can be understood on geometric grounds: when the ideal NiO$_6$ octahedra are 
recovered near 0.5\% strain, the bi-axial lattice distortion 
imposed on the film by the substrate leads to a 
tiling of corner-connected octahedra that is incompatible with unit cell size.\cite{Rondinelli/Coh:2011}

Finally, we note that if the calculations for strained LaNiO$_3$ do not allow for full
relaxation of the out-of-plane lattice parameter, LaNiO$_3$ appears to have the same 
behavior as LaAlO$_3$. This is important for two reasons: First, it indicates that full
structural optimizations are essential, and qualitatively incorrect behavior can be
obtained by artificially neglecting some structural degrees of freedom. And second, 
it points to the strong sensitivity of the octahedral rotations to the details of
the elastic response of the material.

\begin{figure}
\centering
\includegraphics[width=0.45\textwidth]{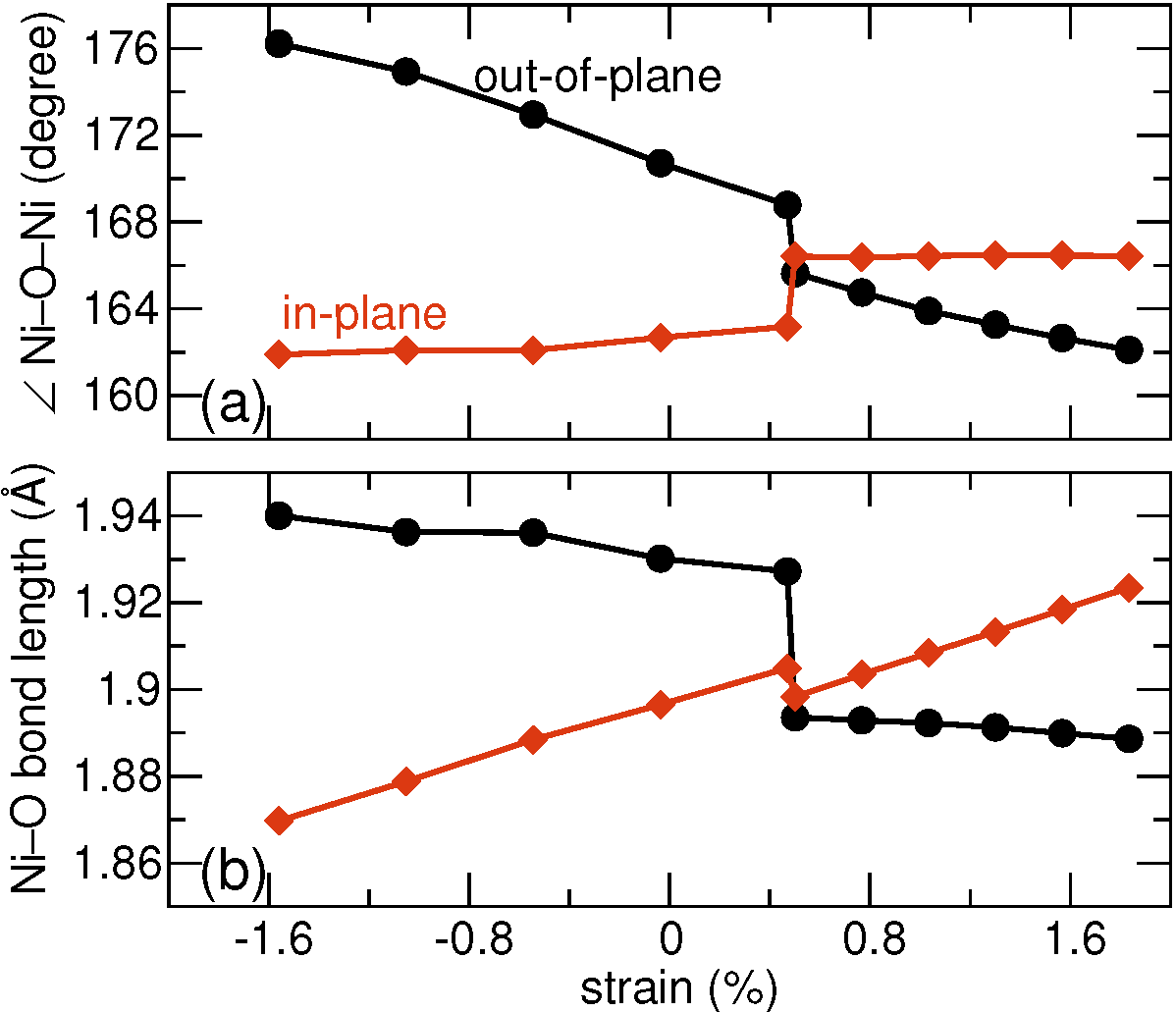}
\caption{\label{fig:lno_data}
The (001)-strain dependence of the in-plane (circles) and out-of-plane (diamonds) 
bond angles and lengths for a homoepitaxial LaNiO$_3$ film.}
\end{figure}

These two examples show that strain clearly couples to the magnitude of the 
octahedral rotations about different crystallographic axes relative to the 
interface. In the next section, we look at how strain influences separately the 
$B-$O$-B$ bond angles and $B-$O bond lengths. 
This separation is important because the effects of bond angle 
and bond length changes on electronic properties, such as bandwidths and
band gaps, and magnetic properties (exchange interactions) are often 
quite different.
Separate control of both parameters, therefore, would be highly desirable
in attempting to engineer specific behaviors.
To address this question, we show in Figure \ref{fig:lno_data} the change in 
in-plane and out-of-plane Ni$-$O$-$Ni bond angles and Ni$-$O bond lengths 
as a function of strain.
The first striking result is that the in-plane bond angle is only weakly sensitive 
to epitaxial strain (0.62$^\circ/$ percent strain, with a discontinuity at the isosymmetric phase
transition), while the out-of-plane angle can be tuned by 3.6$^\circ/$ percent strain.
Conversely, the in-plane Ni$-$O bond lengths are strongly strain dependent, 
since they are accommodating the change in in-plane unit cell area with strain. 
The out-of-plane bond lengths are only weakly strain dependent because changes 
in the out-of-plane \textit{angle} take up the change in out-of-plane 
\textit{lattice parameter}. 
It is clear that a simple picture of strain accommodation via rigid octahedral rotations 
is not appropriate, and in fact changes in bond lengths---in this case specifically
the in-plane bond lengths---mediate a substantial portion of the change in
lattice parameters.


\subsection{Orthorhombic SrRuO$_3$ and CaTiO$_3$}
\begin{figure}[b]
\centering
\includegraphics[width=0.49\textwidth]{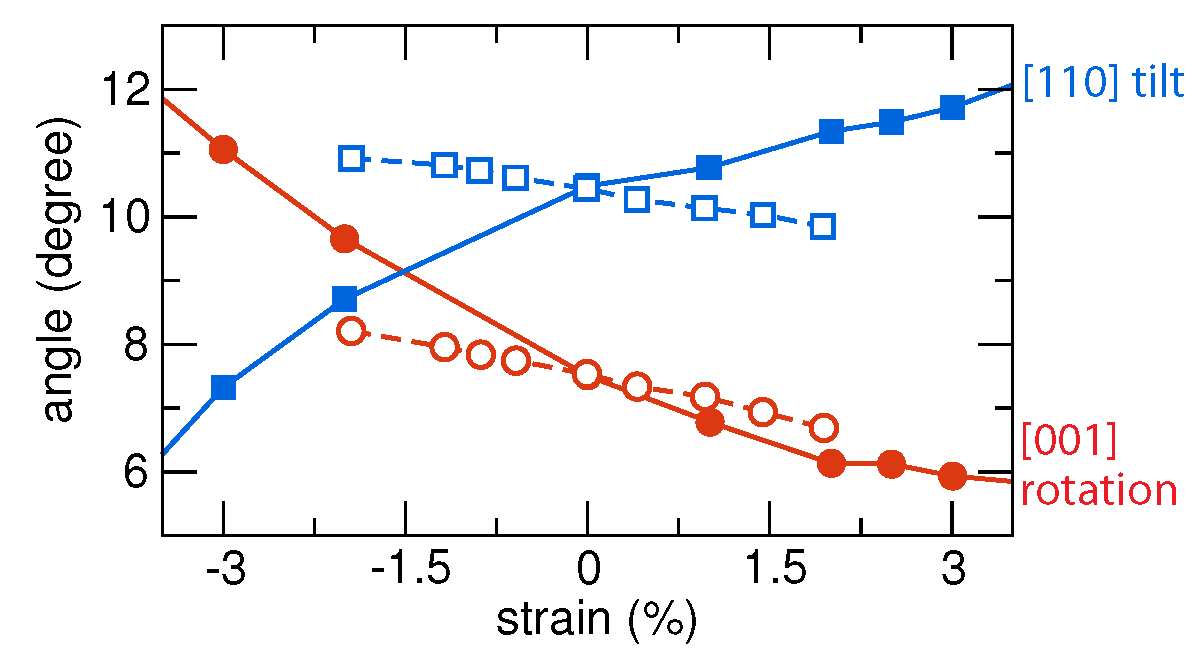}
\caption{\label{fig:sro_summary}
Changes in the RuO$_6$ octahedra rotation angles along the [001]- (circles) and 
[110]- (squares) directions for a homoepitaxially strained SrRuO$_3$ film.
We show the strain dependence for the orthorhombic films with the long 
axis set perpendicular ($c$-oriented, filled symbols) 
and parallel  ($ab$-oriented, open symbols) to the (001)-epitaxial plane.
Data reproduced with permission from Ref.\ ~\onlinecite{Zayak/Rabe:2006}, 
Copyright 2006, {\it American Physical Society}.}
\end{figure}

The ``simple'' rhombohedral perovskites of the previous section were useful 
examples for illustrating the consequences of symmetry lowering and elastic energy 
accommodation in thin films.
Here we discuss two examples -- SrRuO$_3$ and CaTiO$_3$ -- which have the more complex 
orthorhombic $Pnma$ symmetry, with the $a^-a^-c^+$ tilt pattern. This class is particularly
important since it is adopted by the majority of $AB$O$_3$ perovskites.\cite{Woodward:1997b,Thomas:1989}
The thin film behavior of materials with this tilt system is more complicated since the
structure already has orientational anisotropy in the bulk, with the in-phase $c^+$ rotations
along the long axis of the unit cell. As a result the films can have two unique orientations 
on a substrate.\cite{Tagantsev/Cross/Fousek:2010}
When the long axis lies in the epitaxial plane the film is described as
$ab$-oriented and when it is out-of-plane as $c$-oriented.
While experimental thin films often show a mixture of these two 
orientations,\cite{Proffit/Eom:2008,Han/Trolier-McKinstry:2009,Choi/Eom_et_al:2010}
first-principles calculations can determine which is energetically more favorable 
at each strain state, as well as the separate properties of the two orientations. 
In addition, we note that the orthorhombic structure is reached from the cubic
aristotype by the softening of {\it two} zone-boundary instabilities of different symmetry: 
one at the zone corner ($R$-point) and one at the edge ($M$-point). 
In contrast,  the bulk $R\bar{3}c$ structures discussed earlier are 
reached with only one $R$-point instability.
We therefore expect that the $a^-a^-c^+$ tilt pattern might show a quite different
strain response---possibly with less strong coupling between the rotations 
and the strain.
We illustrate the effect of strain on orthorhombic perovskites using two test materials: 
metallic SrRuO$_3$, a common electrode material\cite{Herranz/Fontcuberta_et_al:2004} 
used in thin film growth, and the insulating, 
prototypical perovskite, CaTiO$_3$. 
As in the previous section, one of our choices is a correlated metal,\cite{Cao_et_al:1997,Okamoto/Fujimori:1999,Rondinelli/Spaldin_et_al:2008} 
and the other is a wide-band gap insulator,\cite{Cockayne/Burton:2000} although we will see in this case that the
dielectrically active Ti$^{4+}$ ion in CaTiO$_3$ leads to further complexity compared
with the inert Al$^{3+}$ ion in LaAlO$_3$.
SrRuO$_3$ (CaTiO$_3$) has rotation and tilt angles of 5.8$^\circ$ and 
7.6$^\circ$ (8.7$^\circ$ and 11.7$^\circ$), respectively, with metal--oxygen--metal 
bond angles of 163$^\circ$ (156$^\circ$). \cite{Jones/Harrison:1989,Sasaki/Prewitt_et_al:1987}
Note, these angles are further away from 
180$^\circ$ than our rhombohedral examples, thus 
indicating that the compounds are
more highly distorted.
In Figure \ref{fig:sro_summary}, we show the results of homoepitaxial strain calculations 
by Zayak and co-workers\cite{Zayak/Rabe:2006} for different orientations of orthorhombic 
SrRuO$_3$.
We see that the $a^-a^-c^+$ tilt pattern is maintained for all strain values explored.
For the $c$-oriented structural variant, the in-plane lattice parameters of 
the orthorhombic film are those that set the $a^-a^-$ component of the tilt 
pattern. Since they are already equal in the bulk 
we expect similar behavior to the rhombohedral case.
Indeed, as strain is increased from compressive to tensile, the tilt (rotation) angles 
increase (decrease) as observed previously.
For the $ab$-orientated structure, however, where the lattice parameters of the $c^+$ 
rotation axis and one of the $a^-$ axes are set by the substrate, a markedly different 
strain dependence of the rotation and tilt angle occurs.  
In this case {\it both} tilt and rotation angles decrease (increase) with compressive 
(tensile) strain. 
This is the first example that we have seen of strain giving an overall change in
the magnitude of the angles; in the previous cases a reduction in rotation was always
compensated for by an increase in tilts.
Controlled growth of (110)-oriented orthorhombic perovskites could therefore provide
a route for controlling the $B-$O$-B$ bond angles and as a result the electronic bandwidth 
in functional oxides. 
Evidence of such orientation-dependent electronic properties is reported for 
isostructural LaTiO$_3$ thin films;\cite{Wong/Suzuki_et_al:2010} here, 
LaTiO$_3$ films under nominally similar magnitudes of strain on 
(001)-oriented substrates show robust metallic behavior, while 
those grown on (110)-oriented surfaces are highly insulating.
Zayak and co-workers also calculated the magnetostructural coupling in SrRuO$_3$
and found it to be both substrate orientation  (0.35~$\mu_B$/f.u. difference between
(110)- and (001)-oriented) and strain dependent 
(0.13~$\mu_B$/f.u.\ per percent strain) \cite{Zayak/Rabe:2006}.
Later they predicted an unusual low-spin $S=0$ to high-spin $S=1$ state transition 
for the Ru$^{4+}$ cation under bi-axial elastic strain conditions \cite{Zayak/Rabe:2008}.
While such an on/off control of magnetism with strain in strontium ruthenate 
remains to be confirmed experimentally,\cite{Grutter/Suzuki:2010} the prediction is consistent with 
isovalent chemical substitution studies (that mimic the application of pressure), 
which show that the ferromagnetic ground state is highly susceptible to changes 
in the rotation angles.\cite{Cao_et_al:1997}
Next we describe the calculated strain behavior of orthorhombic $Pnma$ CaTiO$_3$.
A detailed first-principles study by Eklund {\it et al.} \cite{Eklund/Fennie/Rabe:2009}
found similar evolution in the octahedral tilts and rotations as those shown in 
Fig.~\ref{fig:sro_summary} for SrRuO$_3$.
In addition, however, the authors of Ref.\ ~\onlinecite{Eklund/Fennie/Rabe:2009}
found that the relative stability of $ab$- and $c$- orientated structures can 
be tuned with bi-axial strain.
Under compressive and small tensile strains the $ab$-oriented films are more
stable than the $c$-oriented films. However
with increasing tensile strain, the $ab$-oriented films are relatively de-stabilized 
with $c$-oriented films becoming energetically more favorable at around  $+1.5$\% strain. 
Intriguingly, at very large tensile strains (around $+4$~\%), a ferroelectric ground 
state is obtained in the stable $c$-oriented films with the direction of polarization 
along the [110]-direction.
In contrast, no polar phases were found in the $ab$-oriented films between -3\% compressive strain
through the range of stability to 1.5~\% tensile strain.
Because the change in the magnitude of the octahedral rotations under such large 
tensile strains is calculated to be small with respect to the strain-free case, the authors attributed the 
activation of the polar instability to a large strain--polarization coupling term in the 
free energy rather than a competition between polar and rotational instabilities.
We summarize the results up to this point by noting that the octahedral rotations and tilts 
in rhombohedral and $c$-oriented orthorhombic perovskites behave somewhat similarly
in that compressive strain enhances octahedral rotations about the [001]-direction and 
tensile strain favors a [110]-rotation axis with corresponding increases in the tilt angles. 
An exception  occurs in $ab$-oriented orthorhombic perovskites: the magnitude 
of {\it both} angles decreases with increasing tensile strain.
In addition, we have seen in CaTiO$_3$ the first example of strain inducing
a ferroelectric ground state. In the next section, we review further examples 
of such strong strain behavior in materials containing ferroelectrically active ions 
such as Ti$^{4+}$.

\subsection{Oxides with ferroelectrically-active ions and magnetic propensities}
Motivated by the observation in the previous section that strain can induce
ferroelectricity in CaTiO$_3$, we next review additional examples that illustrate
the interaction between strain and ferroelectricity. (For a thorough review 
see Ref.~\onlinecite{Rabe_et_al:2007}.)
First we describe the
behavior of diamagnetic SrTiO$_3$ and ferromagnetic EuTiO$_3$ 
-- two additional cases which are not ferroelectric in the bulk, but in which 
the ferroelectrically active Ti$^{4+}$ ions are ``activated'' by strain. 
\cite{Lin/Guo:2006,Fennie/Rabe:2006},
Next, we show how homoepitaxial first-principles studies discovered that 
perovskites  containing nominally Jahn-Teller \textit{inactive} Mn$^{3+}$
could be coaxed to undergo ferroelectric displacements through 
epitaxial strain constraints.
And finally, we describe the case of BiFeO$_3$, which is already ferroelectric 
and magnetic in its bulk ground state, but where 
strain induces an unusual phase coexistence 
between two structural variants \cite{Hatt/Ramesh:2009}. 
SrTiO$_3$ is not ferroelectric, but is an excellent dielectric\cite{Christen/Mannhart_et_al:1994} 
in which the transition to a ferroelectric state is believed to be suppressed by quantum
fluctuations \cite{Muller/Burkard:1979}.
Instead, the ground state is tetragonal with an $a^0a^0c^-$ tilt pattern of the oxygen 
octahedra. First-principles calculations of bulk SrTiO$_3$ have shown that these
antiferrodistortive rotations compete with and have a tendency to suppress the
ferroelectric instability \cite{Sai/Vanderbilt:2000}.
Early phenomenological studies \cite{Pertsev/Setter:2000} suggested that ferroelectric
polarization could be obtained, and its orientation controlled, by appropriate strain
conditions. 
Subsequent first-principles calculations\cite{Lin/Guo:2006} were consistent with
the phenomenological results.
As in the rhombohedral perovskites, the DFT-based calculations find a change in the 
octahedral rotation axis from [001]$\rightarrow$[110] on going from a compressive to tensile 
strain state.
In addition, polar displacements activated by epitaxial strain are found, with a [110]
orientation of polarization favored by tensile strain, as in the case of CaTiO$_3$.
In contrast to the CaTiO$_3$ case, however, polarization is also induced by compressive
strain, this time along the [001]-direction.  
Only between -0.4\% and +0.2\% strain
is a paraelectric ground state found; in this region the out-of-phase 
rotations of oxygen octahedra dominated the structure.
For larger compressive (or tensile) strains, a ferroelectric polarization is induced 
in the presence of these octahedral rotations.
While ferroelectric hysteresis loops in strained SrTiO$_3$ have not yet been measured
directly, divergence of the dielectric constant, indicative of a ferroelectric phase 
transition, has indeed been reported at room temperature \cite{Haeni/Schlom:2004}.
The ability of first-principles calculations to identify critical strain regions where 
ferroelectric behavior is induced in common dielectrics has spawned numerous 
homoepitaxial strain studies of magnetic dielectrics.\cite{Ramesh/Spaldin:2007,Picozzi/Ederer:2009}
These {\it ab initio} searches for ferroelectricity in magnetic 
materials are motivated by the desire to identify new classes of magnetoelectric 
multiferroics, that is materials with simultaneous and coupled
magnetic and ferroelectric properties.
\cite{Spaldin/Fiebig:2005,Fiebig:2005,Eerenstein/Mathur/Scott:2006,Cheong/Mostovoy:2007}

\begin{figure}
\centering
\includegraphics[width=0.46\textwidth]{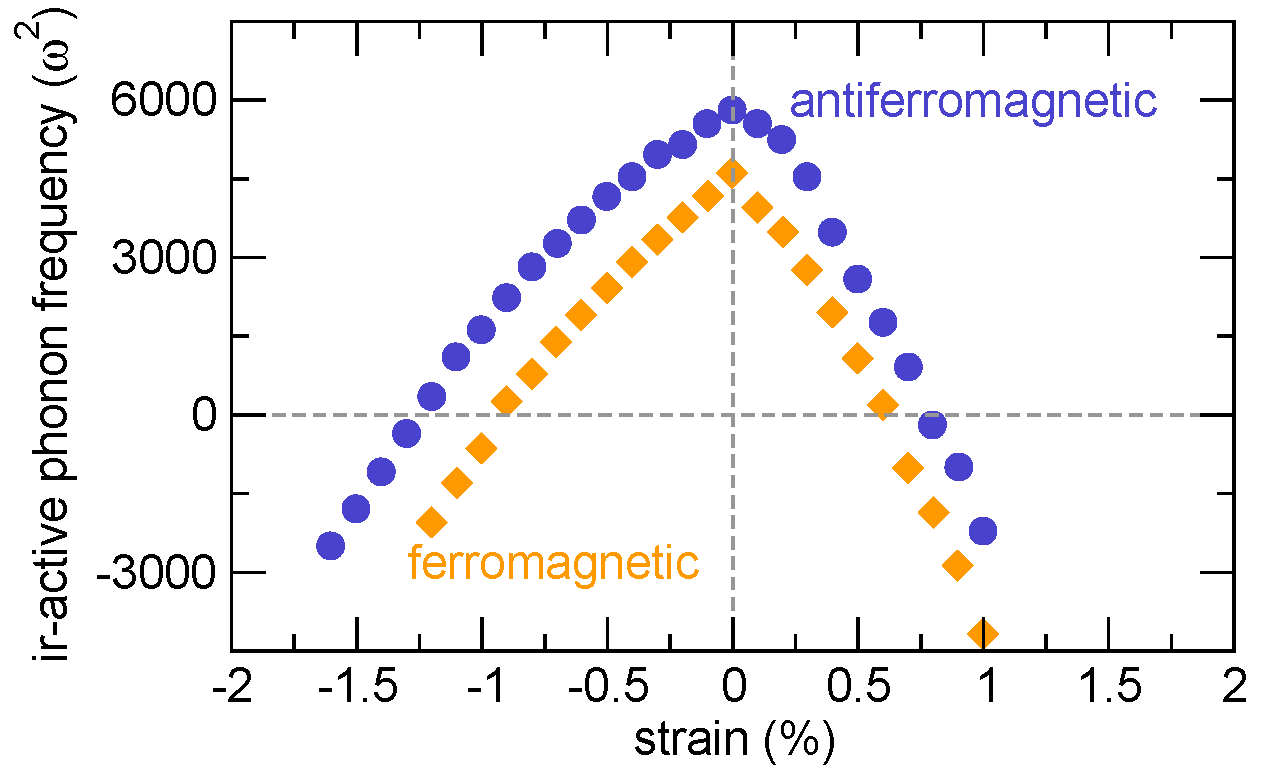}\vspace{-6pt}
\caption{\label{fig:eto_strain}
Evolution of the lowest frequency infrared-active phonon with bi-axial 
strain for cubic EuTiO$_3$ with different magnetic configurations: 
spins aligned parallel (ferromagnetic) and antiparallel (G-type antiferromagnetic).
Strain states for which $\omega^2<0$ indicate a ferroelectric instability.
Data reproduced with permission from Ref.~\onlinecite{Schlom/Fennie_etal:2010}, 
Copyright 2010, {\it Nature Publishing Group}.}
\end{figure}
An example of a material studied with this motivation is perovskite-structure
EuTiO$_3$, which is isovalent with SrTiO$_3$, but has the additional feature
of magnetic $f$-electrons on the Eu$^{2+}$ ions.
Bulk EuTiO$_3$ is reported to have the ideal cubic perovskite structure, with
no ferroelectric polarization, and antiferromagnetic ordering of the local
Eu$^{2+}$ magnetic moments. 
Because the atomic and electronic structures of EuTiO$_3$ closely resemble
those of SrTiO$_3$, a similar strain-induced ferroelectricity should be
anticipated in this case. 
Indeed, homoepitaxial first-principles calculations\cite{Fennie/Rabe:2006} by Fennie and Rabe  
showed that application of $\sim$1~\% compressive strain is sufficient 
to cause the Ti$^{4+}$ cation to off-center in the direction perpendicular 
to the epitaxial plane---similar to the strain-induced ferroelectricity found in SrTiO$_3$.
As in SrTiO$_3$, the strain-induced ferroelectricity in EuTiO$_3$ is 
understood to originate from the strong coupling between strain-induced lattice deformations 
and the lowest frequency transverse optical mode.\cite{Cohen:1992}
The authors of Ref.~\onlinecite{Fennie/Rabe:2006} also found that 
under compressive strain 
the polar mode for a \textit{ferromagnetically} ordered EuTiO$_3$ crystal is of lower 
energy (softer) than that of the \textit{antiferromagnetic} 
spin arrangement at the same strain state (Fig.\ \ref{fig:eto_strain}).
This led Fennie and Rabe to suggest that strain 
could simultaneously modify both magnetic and electric ferroic orders.
%
%
Recent calculations\cite{Schlom/Fennie_etal:2010} showed that an intriguing ferroelectric and 
ferromagnetic EuTiO$_3$ 
phase should also be accessible under tensile strains larger than 0.6\%.
Using those \textit{ab initio} guidelines, 
a stable multiferroic phase, with mutual ferroic coexistence,   
has been subsequently realized experimentally.\cite{Schlom/Fennie_etal:2010}
The perovskite-structure rare-earth and alkaline-earth manganites are of tremendous interest 
because 
of their rich structural, magnetic and electronic phase diagrams and magnetoresistive behavior.
Incorporating ferroelectricity in the insulating members of the series would add another
desirable functionality.
Under usual conditions, however, Mn$^{4+}$ and Mn$^{3+}$ cations do not undergo ferroelectric 
off-centering
because the non-zero $d$-orbital occupation introduces a large electronic penalty for
off-centric distortions.\cite{Hill:2000} 
In this capacity, first-principles calculations have been used
to explore circumstances under which such an off-centering might be induced.
With increasing cation size, the bulk structures evolve from orthorhombic (Ca), with 
large octahedral rotations, to cubic (Sr), and finally a hexagonal structure (Ba), characterized 
by both corner- and edge-shared octahedra.
CaMnO$_3$, although centrosymmetric, in fact has a ferroelectric instability in the
cubic phase that is 
quenched by the $a^-a^-c^+$  octahedral rotations observed in the ground 
state structure.\cite{Khomskii:2006}
Motivated by the sensitivity of the ferroelectric mode to strain, 
Bhattacharjee and co-workers performed first-principles homoepitaxial strain calculations 
for CaMnO$_3$ and identified that the competition between the rotational and ferroelectric 
instabilities favors the polar structure for tensile strains greater than $\sim2$\%. 
\cite{Bhattacharjee/Bousquet/Ghosez:2009}
Here, a ferroelectric polarization develops in the epitaxial plane, 
driven by displacements of the Mn cations\cite{Bhattacharjee/Bousquet/Ghosez:2008} that 
coexists with the competing $a^-a^-c^+$ octahedral rotation pattern.
They also showed that compressive strain (up to 4\%) does not stabilize
the ferroelectric instability. 
Consistent with our earlier discussion, 
the authors of Ref.~\onlinecite{Bhattacharjee/Bousquet/Ghosez:2009} 
found that the frequency of the octahedral rotational instabilities are 
less sensitive to the bi-axial strain than the frequency of the 
lowest polar phonon.
Similar homoepitaxial strain calculations have been performed on perovskite-structured 
SrMnO$_3$,\cite{Lee/Rabe:2010} and BaMnO$_3$.\cite{Rondinelli/Eidelson/Spaldin:2009} 
In SrMnO$_3$, the Mn$^{4+}$ cations undergo ferroelectric off-centering 
under strain, even in the presence of octahedral rotations.
Unlike CaMnO$_3$, however, SrMnO$_3$ shows both out-of-plane (for compressive 
strains larger than 1.4\%) and in-plane (tensile strains larger than 1\%) polarizations.
These critical strains are smaller than that for CaMnO$_3$ 
due to the larger Sr-cation which effectively produces an ``internal'' chemical strain on the 
lattice.
Interestingly, Lee and Rabe also report a  large spin-phonon coupling, similar to that of EuTiO$_3$ 
for SrMnO$_3$: strains greater than approximately $\pm3$\% induce a transition from 
an antiferromagnetic to ferromagnetic spin configuration on the Mn cations which are 
simultaneously displaced from the center of their oxygen coordinating octahedra.\cite{Lee/Rabe:2010}
In hypothetical perovskite-structure BaMnO$_3$,
on the other hand, the larger Ba cation stabilizes the ferroelectric state 
even at its equilibrium volume---no strain is 
required.\cite{Rondinelli/Eidelson/Spaldin:2009}
This occurs because the perovskite phase, which is metastable with respect to the 
denser hexagonal structure, has such a large cell volume that the Mn cation becomes 
severely underbonded. Therefore, the Mn cation off-centers towards the edge
of an octahedron in order to make two strong Mn--O bonds.
The authors of Ref.~\onlinecite{Rondinelli/Eidelson/Spaldin:2009} also show that  the ferroelectric  
perovskite structure becomes the lowest energy phase at very large tensile 
strains, since under those elastic conditions, the denser hexagonal phase is 
energetically unstable.
Ferroelectric behavior has not yet been observed experimentally in any of these 
Mn-based compounds.
We note that similar volume-dependent ferroelectric instabilities have also 
been reported in chromate-based perovskites.\cite{Ederer:2011}
Substantial efforts are underway, however, to explore whether combinations of 
alkaline earth cations  in manganite superlattices subjected to 
various bi-axial strain conditions can stabilize  ferroelectric behavior.

\begin{figure}
\centering
\includegraphics[width=0.45\textwidth]{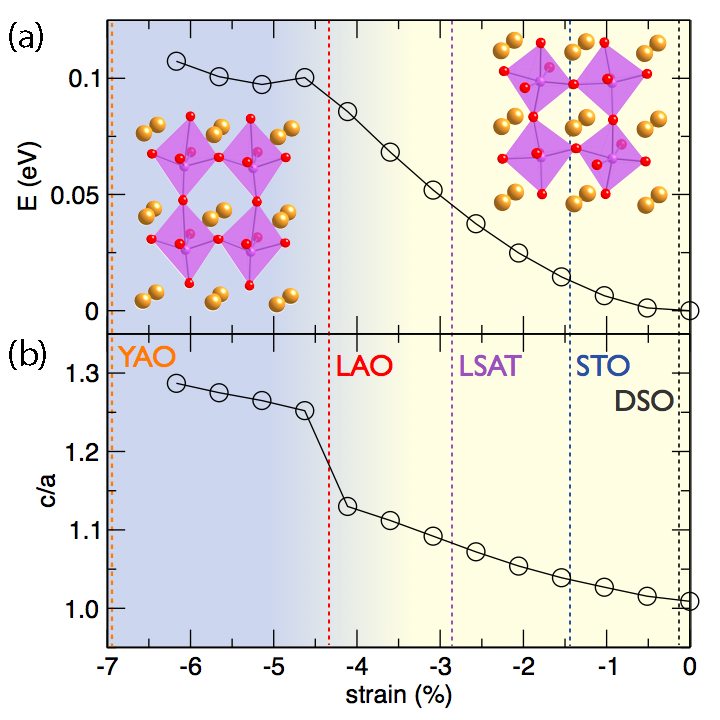}\vspace{-6pt}
\caption{\label{fig:bfo_data}
(a) Evolution in the total energy for BiFeO$_3$ as a function of in-plane strain. 
The insets indicate the two structural variants---both with monoclinic symmetry---with 
the 
long (left) and short (right) $c$-axes that are accessible under large compressive  
and modest strain, respectively.
In (b) the $c/a$ axial ratio shows an abrupt discontinuity at the 
isosymmetric transition near -4.5\% strain. 
The lattice strains corresponding to a number of commonly used oxide 
substrates are shown as dashed lines (see Table~\ref{tab:tilts} for a key to the labels).
Data reproduced with permission from Ref.~\onlinecite{Hatt/Ramesh:2009}, 
Copyright 2009, {\it American Association for the Advancement of Science}.}
\end{figure}
Finally for this section, we discuss the strain-dependence of the behavior in BiFeO$_3$, in which
the  bulk ground state is {\it already} magnetic and ferroelectric, but which shows a strong
evolution of the ferroelectric behavior with strain.
Bulk BiFeO$_3$ has the rhombohedral $R3c$ structure, which consists of 
antiferrodistortive octahedral rotations ($a^-a^-a^-$)around the [111] axis, 
similar to those of LaNiO$_3$ and LaAlO$_3$, and an 
additional relative off-centering of anions and cations along the [111] direction
leading to a ferroelectric polarization along that axis.
In Figure \ref{fig:bfo_data},  
we show the calculated total energy (upper panel) and $c/a$ ratio (lower panel) 
for homoepitaxial BiFeO$_3$ films as a function of in-plane biaxial strain (from Ref.~\onlinecite{Hatt/Ramesh:2009}.
While the symmetry of the system remains monoclinic $Cc$ throughout, there is
an isosymmetric phase transition at $\sim$4\% compressive strain which is characterized
by an abrupt change in $c/a$ ratio, and a change in the coordination environment of
the Fe from [6]-coordinated octahedral to [5]-coordinated square pyramidal.\cite{Hatt/Spaldin/Ederer:2010} 
The transition is accompanied by a re-orientation and enhancement of the ferroelectric 
polarization from $\sim$ 90 $\mu$C/cm$^2$ about an axis close to [111] to 150 $\mu$C/cm$^2$
about an axis close to [001]. 
There is also a change in the octahedral tilt pattern from a 
pattern that is strongly reminiscent of the bulk-like $a^-a^-a^-$ tilt pattern
to rotations mainly about an axis that is close to the out-of-plane direction.%
\cite{Hatt/Spaldin/Ederer:2010}
We note, however, that there are multiple structurally 
unique states (Ref.~\onlinecite{Diguez/Iniguez:2011}),  
which are less than 100~meV/f.u.\ higher in energy that compete 
with this isosymmetric phase transition. If any of these phases
with different symmetry occurs experimentally, the transition is
of course no longer isosymmetric
(see Refs.~\onlinecite{Dupe/Dkhil:2010} 
and~\onlinecite{Spaldin/Christen:2010}).
While the strain value at which this transition is predicted to occur might
be expected to be too large for experimental realization, in fact, thin films of
BiFeO$_3$ grown on YAlO$_3$ (with a 5\% lattice mismatch) do form in the large
$c/a$ structure.
Even more intriguingly, films on LaAlO$_3$, which has a lattice constant 
corresponding to the cross-over point,\cite{Bea/Barthelemy:2009} show a coexistence of the two
phases with a so-called {\it self-morphotropic phase boundary} between
them that can be manipulated by an electric field.
\cite{Hatt/Ramesh:2009}

In summary, we have seen in these examples, first that imposing coherence
with a substrate removes some of the symmetry (diad, triad, or tetrad) 
axes (Table \ref{tab:lattices}) about which the octahedra rotate.
Subsequently, biaxial strain modifies the rotation patterns by altering 
the magnitude of the rotation angles about these axes, and in the extreme
case de-activating or activating new tilt patterns. 
In addition, bi-axial strain deforms the $B$O$_6$ octahedra 
by elongation or compression of the $B$--O bond lengths.
The structural distortion that dominates depends on 
the compressibility of the $B$--O bonds and tendency for the octahedra to 
rotate as gleaned from temperature and pressure experiments on a 
range of perovskites \cite{Zhao/Ross/Angel:2004,Angel/Zhao/Ross:2005}.
The changes in symmetry, bond angles and bond lengths in turn can have profound
effects on the properties of the films. For example, responses that are prohibited
by symmetry in the bulk may become allowed, changes in bandwidths can lead to
drastically different electronic and optical properties, and changes in exchange
interactions can change magnetic properties.
The LaAlO$_3$ and LaNiO$_3$ examples illustrated that, even in ostensible simple
materials, the strain response is in general
complex and simple models such as treating the oxygen octahedra as rigid units, 
or in the opposite extreme ignoring the response of the octahedral rotations may 
not be reliable. 
Materials that are proximal to ferroelectric instabilities, or that are already
ferroelectric in their bulk ground states, show even more complex strain 
responses.
Indeed, simple guidelines for the change in structure with strain are not yet 
available, and until a larger database is established we recommend full first-principles 
calculations with relaxation of all variables rather than models or intuition for 
predicting the structural response of thin film oxides to strain. 

\begin{figure*}
\centering
\includegraphics[width=0.68\textwidth]{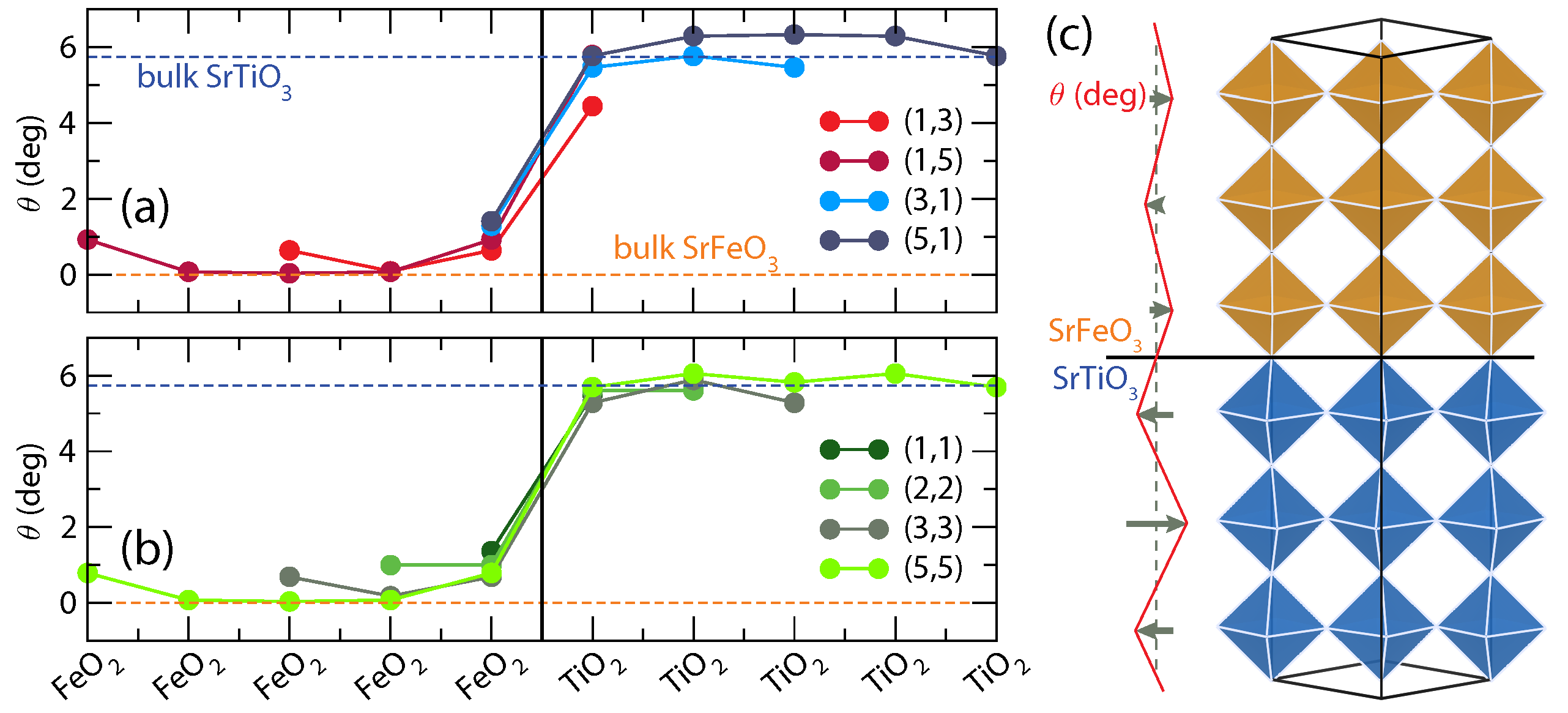}
\caption{\label{fig:sfo_sto} The layer-by-layer resolved octahedral 
rotation angles ($\theta$) for the($n,m$) asymmetric (a) and ($n,m=n$) 
symmetric (b) (SrTiO$_3$)$_n$/(SrFeO$_3$)$_m$ heterostructures. 
The magnitude of the $a^0a^0c^-$ tilt pattern 
rapidly decreases away from the interfacial layers as shown 
schematically in (c) for the (SrFeO$_3$)$_3$/(SrTiO$_3$)$_3$ heterostructure. 
Here, the magnitudes of the octahedral rotations 
about the axis (light, red line) perpendicular to the interface (bold line) are 
indicated by the length of the arrows.
After Ref.~\onlinecite{Rondinelli/Spaldin:2010b}, Copyright 2010, {\it American Physical Society}.}
\end{figure*}
\section{Results: Heteroepitaxial strain.}
It is often asserted in the literature that the symmetry and structure of a substrate
imprint across an interface so that a coherently grown film is affected not only by
the substrate lattice constant but also by the details of its structure. 
In this section, we review the results of electronic structure calculations that have 
been designed to test this hypothesis, with a particular focus on the propagation 
of tilt patterns of oxygen octahedra across interfaces. 
We emphasize again that, in an electronic structure calculation, the effects of 
the presence of an interface can be studied independently from the effects of 
strain by comparing the outcomes of homo- and hetero-epitaxially 
strained systems. 
This provides valuable information, which is difficult, if not impossible to 
obtain experimentally. 

\paragraph{SrFeO$_3$/SrTiO$_3$.} We begin with a study of a ``model'' 
heterostructure system: SrFeO$_3$/SrTiO$_3$ (SFO/STO), which is a 
good prototype system to evaluate the interplay of octahedral rotations 
across a heterointerface for a number of reasons:
First, both materials are formed from neutral (001) planes of $A$O or $B$O$_2$ 
ions, therefore, the heterostructure avoids the complication of a polar 
discontinuity\cite{Hwang_science:2006} (and in turn polar distortions) at 
the heterointerface 
between SrFeO$_3$ and SrTiO$_3$.
Bulk SFO is metallic with p-type conductivity,\cite{Vashuk:2000} and is proximal to multiple 
instabilities: it manifests a long-wavelength spin-density wave, but neither 
Jahn-Teller distorts nor charge orders, even though both possibilities are suggested 
by the high-spin $d^4$ chemistry of the Fe$^{4+}$ ion. 
SrTiO$_3$ is a highly polarizable dielectric, which can couple to 
electronic or structural distortions\cite{Okamoto/Millis/Spaldin:2006} in the SFO layer.
The band alignment across the heterointerface is also Schottky-like within the
LSDA, and the interface does not suffer from pathologies associated with
the DFT underestimation of the band gap\cite{Stengel_et_al:2011}. 
Finally, for the study of octahedral rotations, 
this heterostructure is ideal because the bulk compounds exhibit simple 
oxygen octahedral tilt patterns: SrFeO$_3$ has the ideal cubic 
$Pm\bar{3}m$ perovskite structure ($a^0a^0a^0$ tilt pattern) 
down to the lowest temperature 
studied ($\sim$4~K) \cite{Macchesney/Potter_et_al:1965} and the 
ground state $I4/mcm$ phase of SrTiO$_3$ -- which is a widely used substrate -- 
has a single octahedral instability with 
respect to the cubic phase:\cite{Jauch/Palmer:1999} Below $\sim$105 K it 
exhibits the $a^0a^0c^-$ tilt pattern.
In Ref.~\onlinecite{Rondinelli/Spaldin:2010b}, the authors
investigated the effect of heterostructure periodicity 
in both symmetric (SrTiO$_3$)$_n$/(SrFeO$_3$)$_n$, $n=1\ldots5$, 
and asymmetric  
(SrTiO$_3$)$_n$/(SrFeO$_3$)$_m$, $n=1\ldots3,m=1\ldots3$ superlattices. 
It was found that the octahedral rotations from the SrTiO$_3$ substrate 
propagate into the first two interfacial SrFeO$_3$ layers (Fig.\ \ref{fig:sfo_sto}), 
regardless of the number of SrTiO$_3$ layers. 
The rotational tendencies of the SrTiO$_3$ layers are 
imprinted into the SrFeO$_3$ even in the (1,1) heterostructure.
In heterostructures with ultra-thin (one- or two-layer thick) SrFeO$_3$ layers,
these substrate-induced tiltings combined with the quantum confinement induce 
additional instabilities 
-- charge-ordering and/or Jahn-Teller distortions -- that are not observed in 
bulk SrFeO$_3$.
The authors\cite{Rondinelli/Spaldin:2010b} point out that the octahedral rotations are different for each 
electronic instability: charge ordering prefers the $a^0a^0c^-$ tilt pattern while 
the Jahn-Teller distortions is found to coexist with the $a^-a^-c^+$ tilt pattern.
This \textit{ab initio} result is consistent with a recent group theoretical analysis of octahedral rotations 
and electronically-driven structural distortions.\cite{Carpenter/Howard:2009a,Carpenter/Howard:2009b}
It is worth noting, however, that while specific rotational patterns occur
with each electronic instability it remains unclear whether the rotational
pattern induces the electronic instability or vice versa.
This merits additional study since it offers a possible route to controlling 
electronic phases through octahedral rotations. (A similar suggestion has
also been made recently\cite{Jang/Eom_et_al:2011} for LaTiO$_3$ 
monolayers embedded in SrTiO$_3$ and at manganite/titanate heterointerfaces.
\cite{Segal/Ahn_et_al:2011})
In these highly confined ferrate heterostructures, the Jahn-Teller and 
charge orderings are accompanied by metal--insulator transitions in the 
nominally bulk metallic SrFeO$_3$ layer.\cite{Rondinelli/Spaldin:2010}
Corresponding homoepitaxial strain calculations show that the octahedral 
and electronic lattice instabilities are not induced in SrFeO$_3$ using bi-axial 
strain alone, indicating that substrate coherency and confinement play a critical 
role in determining the structure and properties in these heterostructures.

\paragraph{Manganite Superlattices.} 
A more complicated model system is provided by LaMnO$_3$/SrMnO$_3$ 
superlattices, which combine magnetism with orthorhombic ($a^-a^-c^+$) and 
cubic ($a^0a^0c^0$) symmetries (tilt patterns).
While both constituents are antiferromagnetic insulators, there is 
additional electronic complexity introduced by the $d^4$ Mn$^{3+}$ ion in 
LaMnO$_3$ which has a tendency to Jahn-Teller distortion, and by the polar 
discontinuity -- LaMnO$_3$ has charged (001) layers -- at the interface.\cite{Ederer/Millis:2007}
Both experimental\cite{Yamada/Tokura:2006,Bhattacharya/May_et_al:2008,Aruta/Schlom_et_al:2009}
and theoretical\cite{Rong/Dagotto:2009} studies of these superlattices have 
focused on how the epitaxial strain coupling 
between the spin and orbital degrees of freedom at the heterointerfaces influences the 
macroscopic properties. 
The magnetism and orbital ordering are expected to be highly sensitive to the strain 
condition at the interface, since changes in the bond angles and lengths will alter 
the preferred exchange mechanism and thus the flavor of orbital ordering. 
In addition, possible metallicity due to the polar discontinuity or interfacial mixing 
could change the dominant interactions from super- to double-exchange. 
\begin{figure*}
\centering
\includegraphics[width=0.65\textwidth]{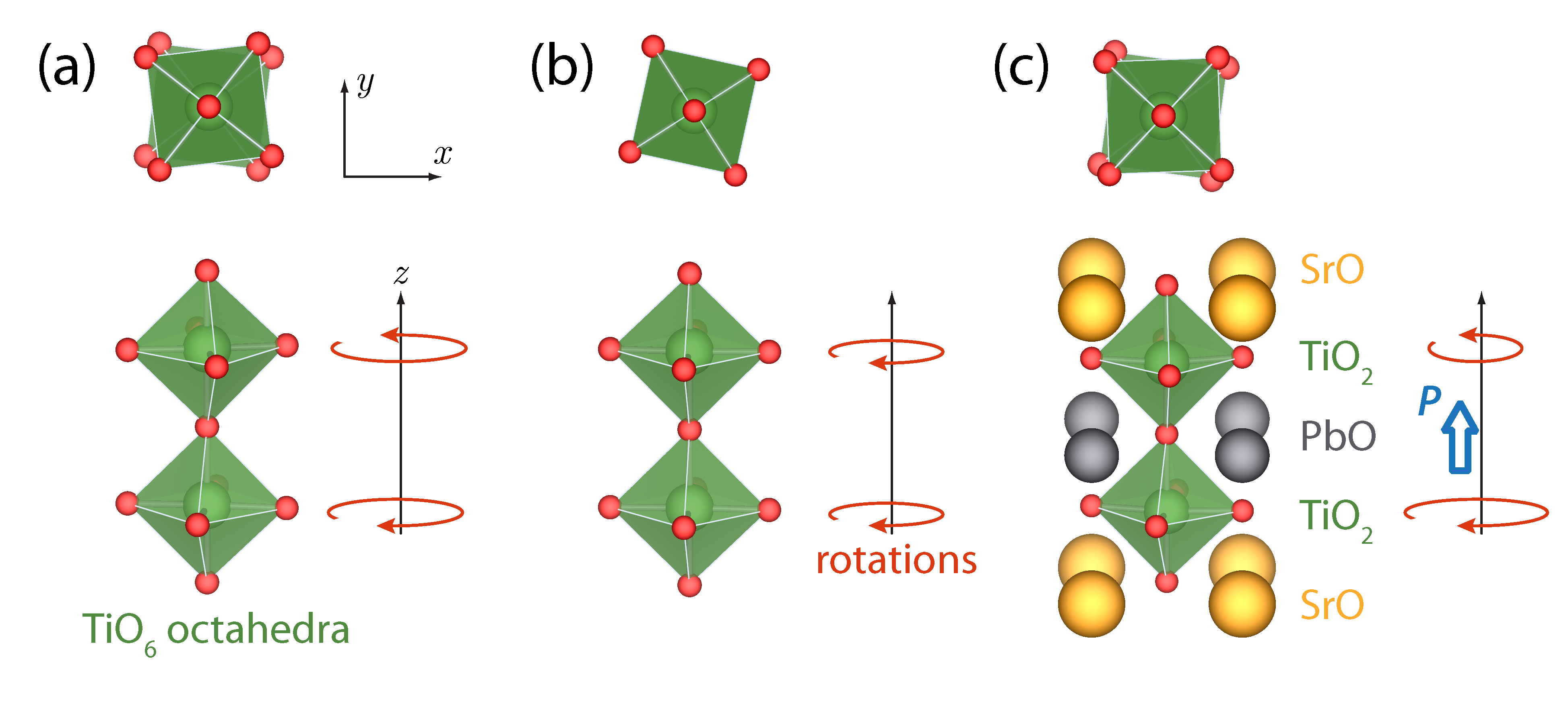}\vspace{-12pt}
\caption{\label{fig:improper}
In ultra-short period (SrTiO$_3$)$_1$/(PbTiO$_3$)$_1$ superlattices 
an unusual combination of out-of-phase (a) and in-phase (b) rotations 
combine about the axis perpendicular to the interface to support a 
ferroelectric polarization along the same direction (c). Here, the arrows indicate 
the rotation directions of the octahedra in each layer and their magnitude 
corresponds to their relative amplitude.
After Ref.~\onlinecite{Bousquet/Ghosez_et_al:2008}, Copyright 2008, {\it Nature 
Publishing Group}.}
\end{figure*}

Experimentally it has been found that the orbital degree of freedom is indeed  
strongly modulated by the strain state \cite{Yamada/Tokura:2006}: 
In compression a $C$-type insulating antiferromagnetic state is stabilized, while 
in contrast, tensile strain produces an $A$-type conducting interface.
For the lattice-matched case, ferromagnetic order is observed, consistent 
with a disordered orbital state.
First principles-calculations of layer-resolved band structures 
\cite{Nanda/Satpathy:2009} indicate a crystal field degeneracy 
splitting induced by the interface strain that supports this 
interpretation of the above experiments; in addition a 
spin-polarized electron gas is calculated to occur at the interface 
in larger period manganite superlattices due to polar mismatch effects
\cite{Nanda/Satpathy:2008b}.
However, we emphasize that the calculations of Refs.~\onlinecite{Nanda/Satpathy:2009}
and~\onlinecite{Nanda/Satpathy:2008b} did not allow for the presence of rotations 
or tilts of the oxygen octahedra, which we have seen can drastically alter the
physics. In fact, modulations in the octahedral rotations in such manganite superlattices 
have been shown experimentally to alter the magnetic ordering 
temperature\cite{May/Bhattacharya:2009}---effects of local octahedral 
distortions in artificial structured materials are thus an obvious area for future study.
There have been a number of recent experimental measurements of octahedral rotations
across interfaces, that are broadly consistent with the picture that is emerging
from the first-principles calculations. For example
real space mapping\cite{Borisevich/Ramesh_et_al:2010} of the octahedral rotations across the 
La$_{0.7}$Sr$_{0.3}$MnO$_3$/BiFeO$_3$ (LSMO/BFO) 
interface with scanning transmission electron microscopy show 
the rotations propagating across the interface although modulated in magnitude
from their bulk values to avoid energetically costly frustrations of the rotations 
across the heterointerface. 
And in (LaNiO$_3$)$_{n}$/(SrMnO$_3$)$_m$ superlattices, the penetration length 
of the rotations across the interface has been 
shown to depend on the distance between different perovskite layers 
composed of large and small rotation angles.\cite{May_et_al:2011}

\paragraph{More exotic behaviors -- some consequences of symmetry lowering.}
Finally, we briefly mention some exotic behaviors in which the symmetry lowering
associated with the presence of a heterointerface allows new properties to 
develop that are genuine properties of the interfacial system rather 
than either bulk parent compound. 
First-principles calculations were recently used to demonstrate a novel kind of
{\it improper ferroelectricity} -- in which the primary order parameter for 
the phase transition is not the ferroelectric polarization -- 
1/1 period superlattices 
of ferroelectric/paraelectric PbTiO$_3$/SrTiO$_3$.\cite{Bousquet/Ghosez_et_al:2008}
In this case, tetragonal ferroelectric PbTiO$_3$ (which does not have 
octahedral rotations) is combined with antiferrodistortive SrTiO$_3$ and 
an enhanced polarization is obtained; the conventional bulk description of 
ferroelectricity would suggest that interruption of the cooperative 
Ti displacments in the PbTiO$_3$ layers by paraelectric SrTiO$_3$ 
should attenuate the macroscopic polarization.
The authors of Ref.~\onlinecite{Bousquet/Ghosez_et_al:2008} 
showed that because of the competition among the structural instabilities at 
the heterointerface,  an unusual antiferrodistortive rotation of the oxygen 
octahedra about an axis perpendicular to the interface is stabilized. This
is symmetry-compatible with 
a ferroelectric polarization along that direction (Fig.\ \ref{fig:improper}). 
Bousquet and co-workers demonstrated theoretically that the enhanced polarization 
is driven by the specific octahedral rotation pattern that is present at the 
heterointerface but absent in the bulk constituents.
Motivated by the predictions, thin films of PbTiO$_3$/SrTiO$_3$ were grown
and a very large and temperature-independent dielectric constant, typical of 
improper ferroelectrics but unusual for conventional ferroelectrics, was measured.\cite{Bousquet/Ghosez_et_al:2008} 
A similar enhancement in interfacial ferroelectric polarization was also predicted 
using heteroepitaxial DFT-based calculations for 
asymmetric and symmetric superlattices of paraelectric CaTiO$_3$ 
and ferroelectric BaTiO$_3$.\cite{Wu/Vanderbilt:2011}
Here, Wu \textit{et al.} found that  large TiO$_6$ octahedra rotations 
persist in all superlattices studied when the adjacent oxide layers are CaO, 
but the amplitude of the rotations 
is substantially reduced when the adjacent layers are BaO. This behavior 
is consistent with the bulk structure of CaTiO$_3$ (BaTiO$_3$) 
which has the  $a^-a^-c^+$ ($a^0a^0c^0$) tilt pattern. 
With an increasing ratio of CaO to BaO layers, or vice versa, the 
rotation magnitudes approach their respective bulk values;  
however, in the ultra-short 1/1 limit the rotations are about half the 
size (4$^\circ$ - 6$^\circ$) of those in bulk CaTiO$_3$.\cite{Wu/Vanderbilt:2011}
Because the octahedral rotations and ferroelectric displacements compete with 
each other in bulk CaTiO$_3$, the
suppressed octahedral rotations at the heterointerface 
between BaO and CaO layers allow for a larger polarization to develop in that layer, 
which in turn enhances the net polarization of the superlattice.
Since these structural distortions strongly couple to bi-axial strain, 
they could be further enhanced by growth on a suitable substrate.
In the previous two examples, the oxide layers 
at the heterointerface break inversion symmetry in the synthetic perovskite 
because the chemical and structure environments in directions perpendicular 
to the interface are inequivalent.
Since the linear magnetoeletric effect can only be non-zero in the absence of 
time-reversal and space-inversion symmetries,  heterointerfaces can be used to
enable magnetoelectric response in otherwise centrosymmetric magnetic materials.
In Ref.~\onlinecite{Rondinelli/Stengel/Spaldin:2008}, first-principles calculations
were used to demonstrate such a linear magnetoelectric effect at the interfaces
in SrRuO$_3$/SrTiO$_3$ superlattices. The effect is symmetry prohibited in 
both parent compounds, but is allowed at the interface.
The authors demonstrated that the magnetoelectric
response arises from a carrier-mediated mechanism, and should be a 
universal feature of the interface between a dielectric and a spin-polarized metal; 
it has subsequently been confirmed in a CoPd film immersed in
an electrolyte\cite{Zhernenkov_et_al:2010} and all-solid-state ferromagnetic 
(La,Sr)MnO$_3$/Pb(Zr,Ti)O$_3$ interface.\cite{Molegraaf/Ahn/Triscone:2009,Vaz/Walker_et_al:2010}
Similar calculations have now been performed on a range of ferroelectric/ferromagnetic 
metal interfaces and novel interfacial multiferroic behavior reported (see for example 
Refs.~\onlinecite{Duan/Jaswal/Tsymbal:2006,Duan/Tsymbal:2008,Cai/Sai_et_al:2009,Yamauchi/Sanyal/Picozzi:2007}
It is important to note however, that calculations for ferroelectric/metal interfaces
are fraught with technical difficulty, because the DFT underestimation of the electronic 
band gap in the insulator can lead to calculated ohmic contacts in 
situations where a Schottky barrier occurs experimentally. 
As a result, spurious real-space charge transfer occurs and this can obscure 
the intrinsic behavior of perovskite heterointerfaces with competing 
structural and electronic instabilities.
For a detailed discussion of the unphysical behaviors caused by this pathology, including 
many examples from the existing literature, see Ref.~\onlinecite{Stengel_et_al:2011}.
Methods such as the recently developed formalism for performing density functional
calculations for capacitors with constrained values of the dielectric displacement
\cite{Stengel/Vanderbilt/Spaldin:2009b,Stengel/Vanderbilt/Spaldin:2009c} 
go some way towards alleviating this problem.
\section{Outlook for Rational Oxide Heterostructure Design\label{sec:outlook}}
In this review we described how first-principles calculations based on density 
functional theory can be used to isolate atomic and electronic structure changes 
in perovskite oxide thin films and heterointerfaces.
We described efforts using the homoepitaxial and heteroepitaxial strain 
approaches to decouple the intrinsic contributions that epitaxial strain, changes 
in symmetry, and interface chemistry play in determining the properties of oxide 
heterostructures.
Several new ideas for oxide heterostructure and thin film design emerge from this
review: 
\begin{itemize}
\item[$\diamond$] The macroscopic properties of oxide heterostructures can often not be 
simply predicted from consideration of the electronic structure of the 
bulk materials alone; the interfacial physical, electronic and magnetic structures
in artificial geometries can be genuinely different from those of the parent bulk 
materials due to changes in symmetry- and size-dependent properties. 
\item[$\diamond$]  Bi-axial strain does more than just change bond lengths
in perovskite thin films; it can couple to and/or alter the internal 
degrees of freedom. In particular, octahedral rotation patterns are modulated 
by strain in a fashion that is not immediately intuitive and can give rise to 
new electronic states.
\item[$\diamond$]  The heterostructure ground state is often influenced by latent instabilities 
present in the bulk phases; the substrate-induced heteroepitaxial constraints 
then act to enhance any unstable modes and to re-normalize the low energy 
electronic structure.
\item[$\diamond$] 
Translational and point symmetry changes at a heterointerface can lift 
bulk electronic degeneracies to promote new and allowed 
order parameter couplings, through for example, strain-induced 
crystal field splittings, which in turn strongly affect the macroscopic behavior.
\end{itemize}

{\noindent \bf Future research directions.$\quad$}
Finally we outline some on-going and new research directions and pressing
open questions in the field that we find of particular interest.
\begin{enumerate}
\item {\it Mechanisms for strain accommodation.} 
We have seen that structurally similar perovskite oxides behave differently under epitaxial strain 
both from each other, and from their bulk counterparts under hydrostatic pressure.
While much of the elastic strain accommodation in heteroepitaxy is accommodated through
changes in both the bond lengths and octahedral rotation patterns, it is still 
unclear why some oxides show relatively larger bond length changes, while others undergo 
larger changes in the octahedral rotation and tilt patterns.
The answer likely is found in the different compressibility of certain transition metal
-- oxygen bonds; future work should attempt to quantify this by surveying a variety of 
insulating and metallic perovskite oxides, and analyzing the stiffnesses of octahedral 
rotations and bond length distortions.
By collecting these data, it may be possible to build a set of design 
rules governing the tendency for certain classes of perovskite oxides 
to undergo different atomic displacement patterns with bi-axial strain.

\item {\it Octahedral texturing.} 
We have seen many examples in which bi-axial strain strongly modifies the octahedral rotation 
patterns in thin films by altering the symmetry axes about which the octahedra rotate.
In all cases, when the in-plane lattice parameters are equal and the transition 
metal nearest neighbor distances are the same, then
the rotations around the $x$- and $y$-axes are equal (or equivalently,
the net rotation is about the [110]-direction).
However, when the two in-plane lattice parameters differ, this degeneracy
is lifted and different directions of rotation are adopted\cite{Hatt/Spaldin:2010}. 
This geometric constraint suggests that the substrate miscut angle -- which 
can change the effective in-plane $a$ and $b$ lattice parameters adopted by a coherent 
film -- could
be a useful parameter to control the octahedral rotations and the 
preferred {\it rotational easy axis}.
Such crystallographic substrate-tailoring could make it possible to achieve 
monodomain samples, or obtain a specific number of 
antiphase domain boundaries.
It has recently been demonstrated experimentally that substrate orientation can 
be used to modify the orbital ordering patterns in manganite compounds: a 
``striped'' phase can be transformed to 
a ``checkerboard'' phase by switching from growth on a (001) to 
(110) terminated surface \cite{Okuyama/Tokura_et_al:2009}. 
The two surface terminations in this case reflect the extreme situation of 
a substrate miscut angle, and we propose that by slowly tuning that angle, 
both orbital polarization and octahedral rotation orientation might be
controllable in related thin film oxides.
\item{\it Intrinsic defect profiles.} 
An important question, which in principle is accessible through first-principles
calculations, is whether the intrinsic defect profile of a film changes with
strain or heteroepitaxy. 
It is widely believed in the oxides community that the lattice constant of 
perovskite oxides depends on the concentration and type of intrinsic defects. 
\cite{Tarsa_et_al:1996}
Indeed, accurate measurements of lattice constants have 
sometimes been used to infer oxygen vacancy 
concentrations.\cite{Brooks/Schlom_et_al:2009,Mitchell/Bader_et_al:1996}
Therefore, it is likely that at large strain values, it may become more energetically
favorable to accommodate changes in lattice constants through changes in the defect 
profile rather than in changes of the bond lengths and tilt angles---a chemical 
strain relaxation mechanism\cite{Selbach/Grande_et_al:2011} consistent with  
Vegard's law.\cite{Vegard:1921,Denton/Ashcroft:1991}
In addition to being of fundamental interest, this issue is of profound technological 
importance since the properties of oxide films often depend sensitively on defect
concentrations.
While in an experiment, multiple defects -- for example oxygen vacancies and cation
non-stoichiometry -- usually occur simultaneously, in calculations it is possible
to separately evaluate the effect of each individual defect on the lattice parameters
of the system. 
A comprehensive series of such calculations would be helpful in identifying
the intrinsic changes in lattice parameters with defect profile, and in turn
likely changes in defect profile with strain.
In practice, however, calculations for realistic defect profiles require large 
supercells, and accurate calculations of stresses induced by the introduction
of defects require large energy cut-offs. 
Therefore such a study is a formidable task for future work.

\item{\it Technical issues.}
It is a well-known problem within density functional theory that standard 
exchange-correlation functionals such as the local density approximation 
have errors in their calculation of lattice constants of up to a few
percent. 
These errors can in turn have rather drastic consequences on the properties.
For example, the prototypical ferroelectric BaTiO$_3$ is paraelectric
at the theoretical local density approximation (LDA) lattice constant, while 
paraelectric SrTiO$_3$ is ferroelectric at the 
theoretical GGA 
lattice constant.\cite{Nishimatzu/Waghmare_et_al:2010}
In standard ``bulk'' calculations, this is often circumvented by working
at the experimentally measured lattice constants. 
This is not a possibility, however, when strain is to be used as a variable
in a calculation, as the experimental lattice parameters have a theoretical 
strain associated with them, and the only well-defined zero-strain reference
state is the calculated structure.  
Development of exchange-correlation functionals that yield accurate bulk
lattice parameters and structures, and testing of their behavior in strained systems,
is therefore a crucial direction for future research.
Here, the recently introduced GGA exchange-correlation functional 
PBEsol\cite{PBEsol:2008}, which is 
biased towards more accurately reproducing surface energies and 
lattice constants than its PBE (Perdew-Burke-Ernzerhof) predecessor\cite{Perdew/Burke/Ernzerhof:1996},  
seems to be particularly promising.\cite{Haas/Tran/Blaha:2009}
\end{enumerate}
In summary, first-principles studies have shown that prior conventional wisdom 
guiding TMO heterostructure design as simple two-component composites should 
be re-evaluated.
Instead, understanding the electronic phases at oxide heterointerfaces
requires self-consistent treatment of the electronic and atomic degrees of 
freedom of both constituents on an equal footing.
We have seen many examples in which, because of the many competing 
electronic and structural degrees of freedom in perovskite oxides, the 
physical properties found in oxide--oxide heterostructures are highly 
susceptible to subtle changes in elastic strain and dimensionality. 
This critical understanding of how correlated electron and 
emergent behavior develops from changes in \textit{local} structure 
and artificial geometries is essential to engineering their functionality.
%

\bibliography{rondo}
\end{document}